\documentclass{article}
\usepackage[T1]{fontenc}
\usepackage{systemind_research,times}
\usepackage{amsmath,amsfonts,bm}

\def\eqref#1{equation~\ref{#1}}
\def\1{\bm{1}}

\DeclareMathAlphabet{\mathsfit}{\encodingdefault}{\sfdefault}{m}{sl}
\SetMathAlphabet{\mathsfit}{bold}{\encodingdefault}{\sfdefault}{bx}{n}

\usepackage{hyperref}
\usepackage{longtable}
\usepackage{url}
\usepackage{booktabs}
\usepackage{multirow}
\usepackage{array}
\usepackage{float}
\usepackage{amsmath,amssymb}
\usepackage{xcolor}
\usepackage{graphicx}
\usepackage{tikz}
\usepackage{xspace}
\usepackage{tabularx}
\usepackage[most]{tcolorbox}
\definecolor{traceBlue}{HTML}{17365D}
\definecolor{traceGold}{HTML}{B99545}
\definecolor{sourcePurple}{HTML}{75618F}
\newtcolorbox{sourcebox}{enhanced,breakable,colback=sourcePurple!5!white,
  colframe=sourcePurple!65!white,boxrule=0.5pt,arc=2pt,
  left=6pt,right=6pt,top=4pt,bottom=4pt,fontupper=\small,
  before skip=6pt,after skip=6pt}
\newtcolorbox{tracebox}[1]{enhanced,breakable,colback=traceBlue!3!white,
  colframe=traceGold,colbacktitle=traceBlue,coltitle=white,
  fonttitle=\bfseries,title={#1},boxrule=0.7pt,arc=2pt,
  left=8pt,right=8pt,top=6pt,bottom=6pt,before skip=10pt,after skip=10pt}
\usetikzlibrary{arrows.meta,positioning,fit,calc}

\newcommand{\rac}{\textsc{RAC}\xspace}
\definecolor{taskGainStrong}{HTML}{084522}
\definecolor{taskDecline}{HTML}{8B0000}
\newcommand{\taskscore}[2]{%
  \ifdim #2pt<-3pt
    \textcolor{taskDecline}{\mbox{#1\,[#2]}}%
  \else\ifdim #2pt>3pt
    \textcolor{taskGainStrong}{\mbox{\textbf{#1\,[#2]}}}%
  \else
    \mbox{#1\,[#2]}%
  \fi\fi}

\title{Can AI Scientists Coordinate at Runtime?}

\author{\parbox{\dimexpr\textwidth-2\tabcolsep\relax}{\raggedright\normalfont
Zijian Liu\textsuperscript{1}, Yangzhixin Luo\textsuperscript{2}, Junyu Lu\textsuperscript{3}, Yi Li\textsuperscript{1}, Yu Chen\textsuperscript{1}, David Xu\textsuperscript{1},\\
William F. Shen\textsuperscript{4,*}, Xinchi Qiu\textsuperscript{4,*}, Xisen Wang\textsuperscript{1,\ensuremath{\dagger}}\\[0.5em]
{\small
\textsuperscript{1}University of Oxford (ox.ac.uk)\\
\textsuperscript{2}King Abdullah University of Science and Technology (kaust.edu.sa)\\
\textsuperscript{3}University of Sydney (usyd.edu.au)\\
\textsuperscript{4}University of Cambridge (cam.ac.uk)}
}}

\systemindfinalcopy
\begin{document}
\maketitle
\begingroup
\renewcommand{\thefootnote}{*}
\footnotetext[1]{William F. Shen and Xinchi Qiu serve only in an advisory role.}
\renewcommand{\thefootnote}{\ensuremath{\dagger}}
\footnotetext[2]{Corresponding author. Contact: \href{mailto:research@systemind.io}{\texttt{research@systemind.io}}}
\endgroup

\begin{abstract}
Multi-agent AI scientists have shown improving performance across a diverse range of tasks. Yet a common approach is design-time agentic orchestration, which typically relies on fixed workflows. In contrast, human scientists coordinate and adjust their division of labor at runtime. We therefore ask: can AI scientists also coordinate at runtime? To this end, we introduce Runtime Agent Coordination (\rac), which selects agents from existing AI-scientist hosts during execution, assigns scoped work contracts, and provides artifact-grounded verification. Verification informs subsequent agents without blocking transitions or discarding artifacts. We conduct a single-seed exploratory evaluation across Agent Laboratory, EvoScientist, and ARK on ResearchClawBench, preserving host models, tools, and permissions under host-calibrated budgets. Four cumulative conditions separate native execution, runtime communication, runtime selection, and the combined addition of contracts and verification. Runtime selection yields the highest observed mean score for each host; adding contracts and verification reduces these means, with host-dependent outcomes relative to native execution. These results motivate runtime coordination while exposing the limits of additional coordination mechanisms under constrained budgets. Code is available at \url{https://github.com/systemind-team/Runtime-AI-Scientist}.
\end{abstract}

\section{Introduction}

AI scientists automate increasingly large parts of the research lifecycle, including idea generation, literature synthesis, experiment design, implementation, analysis, and paper writing~\citep{aiscientist2024,aiscientistv2,agentlab2025,goel2025training, coscientist2025,aris2026}. Multi-agent systems distribute these activities across specialized roles that exchange prompts, critiques, code, and research artifacts. This communication determines what agents share. Coordination is a separate decision: which agent should act next, what it should produce, whether its output is valid, and how the run should continue.

Many prominent AI-scientist systems organize agent interactions through predefined stages, loops, and fallback rules~\citep{aiscientist2024,aiscientistv2,airesearcher2025,agentlab2025,xin2026ark}. Search, repair, and debate add flexibility within these workflows, but do not necessarily let the system select the most relevant agent across stages~\citep{aiscientistv2,autoresearchclaw2026}. Other systems, such as EvoScientist~\citep{evoscientist2026}, incorporate persistent memory and self-evolution. These differences motivate us to investigate whether a shared runtime coordination approach can improve research performance across hosts with varying degrees of native flexibility. 

Selecting the next agent based on the current research state matters because long-horizon scientific failures are state dependent. A missing result calls for an experimenter, a broken bibliography calls for a deterministic check or writer, or a contradictory figure may require code rather than another prose revision. Recent evaluations locate substantial AI-scientist failures in experimental design, grounding, and scientific reasoning rather than only in software crashes~\citep{researchclawbench2026,autoresearchfail2026}. A fixed workflow cannot remove these failures merely by adding more agents. It must either execute all contingencies, which increases cost, or omit contingencies that later turn out to be necessary.

\begin{figure}[H]
\centering
\includegraphics[width=\linewidth]{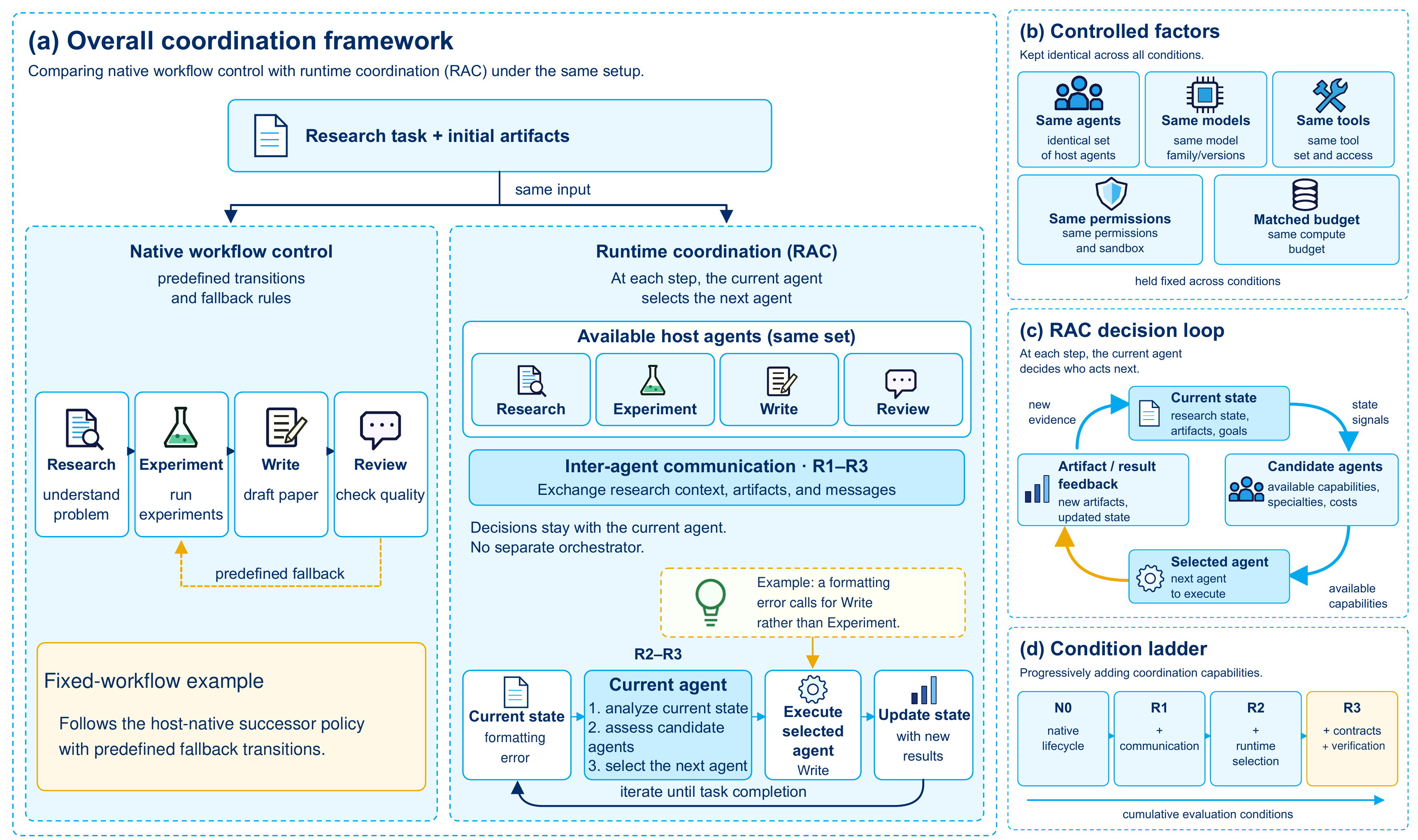}
\caption{\textbf{Same agents, different coordination.} (a) Native transitions versus state-dependent selection by the current agent, with inter-agent communication for exchanging context, artifacts, and messages; this schematic does not imply that all hosts lack runtime delegation. (b) Agents, models, tools, permissions, and run-level budget are held fixed. (c) At each step, the current agent selects the next available agent based on current state, without a separate orchestrator; returned artifacts update the state for the next decision. (d) The cumulative conditions add communication (R1), runtime selection (R2), and contracts plus verification (R3) to the native lifecycle (N0).}
\label{fig:intro_overview}
\end{figure}

We therefore ask whether runtime coordination is a useful alternative to fixed workflows for multi-agent AI scientists~\citep{wang2026hardcoding}. To address this question, we propose Runtime Agent Coordination (\rac), illustrated in Figure~\ref{fig:intro_overview}, which uses current artifacts, open problems, execution history, and remaining budget to select an authorized existing agent. Scoped contracts define what each selected agent should accomplish, while artifact-grounded verification provides feedback on its output for subsequent work. Every verdict retains the workspace and host transition. We apply the same framework across different hosts under matched budgets, keeping their underlying agents and tools fixed to assess the benefits of communication and coordination.

Our core contributions are threefold. First, we formally study runtime coordination in multi-agent AI scientists and formulate next-agent selection as a state-dependent decision, distinguishing it from communication under native workflow control. Second, we instantiate the \rac framework in multi-agent AI scientists such as Agent Laboratory, EvoScientist, and ARK without replacing their native agents, tools, or permissions. Third, we empirically test whether runtime coordination improves research performance under matched budgets, reporting research scores, token usage, and estimated inference costs across a range of tasks.

\section{Related Work}

\subsection{AI Scientists}

AI scientists aim to automate research activities ranging from idea generation and experimentation to analysis and scientific writing~\citep{aiscientist2024,agentlab2025,airesearcher2025}. The AI Scientist organizes these activities into an end-to-end workflow, while AI Scientist-v2 introduces agentic tree search within experimentation~\citep{aiscientist2024,aiscientistv2}. Agent Laboratory assigns specialized agents to research phases, AI-Researcher combines literature exploration with experimental development, and ARK provides a human-steered research harness~\citep{agentlab2025,airesearcher2025,xin2026ark}. These systems demonstrate different ways to organize specialized capabilities across the research lifecycle.

Other systems emphasize iterative feedback and adaptation. Co-scientist combines hypothesis generation, debate, and ranking~\citep{coscientist2025}; AutoResearchClaw integrates multi-agent debate, execution repair, and verifiable reporting~\citep{autoresearchclaw2026}; and ARIS uses cross-model review and evidence checks to assess research claims~\citep{aris2026}. EvoScientist accumulates reusable research experience through persistent memory and self-evolution~\citep{evoscientist2026}. Moving beyond predefined pipelines, freephdlabor supports workflows determined by real-time agent reasoning, while AutoScientists organizes decentralized teams around evolving experimental evidence~\citep{freephdlabor2025,autoscientists2026}. Our work examines whether a shared runtime coordination approach can improve existing hosts with different native control mechanisms.

Scientific output also requires evaluation beyond successful execution. ResearchClawBench evaluates end-to-end research on real-paper-derived tasks, while diagnostic studies examine failures in experimental design, grounding, and scientific reasoning~\citep{researchclawbench2026,autoresearchfail2026}. Automated review introduces further uncertainty because assessments can vary across evaluator models~\citep{canaievaluate2026}. We therefore distinguish feedback used during research from external benchmark evaluation of the final outputs.

\subsection{Multi-Agent Coordination}

Multi-agent systems combine specialized roles through communication and structured collaboration. AutoGen supports programmable agent conversations, MetaGPT combines role-specific workflows with shared structured messages, and multi-agent debate iteratively exchanges candidate answers and critiques~\citep{autogen2023,metagpt2023,du2023debate}. These mechanisms shape the information available to agents, while coordination determines how their contributions are organized and which agent acts next.

Several approaches explicitly optimize or adapt this organization. GPTSwarm represents agents as computational graphs and optimizes prompts and inter-agent connectivity~\citep{gptswarm2024}. DyLAN selects agent teams and supports dynamic interactions during task solving~\citep{dylan2023}. In scientific applications, freephdlabor and AutoScientists further demonstrate reasoning-driven workflows and self-organizing collaboration~\citep{freephdlabor2025,autoscientists2026}. Building on these directions, \rac studies runtime selection across existing AI-scientist hosts while preserving their underlying agents, tools, and permissions. Its cumulative conditions distinguish communication under native control from runtime selection and the combined addition of contracts and verification.

Feedback and resource use are integral to this comparison. MetaGPT incorporates executable feedback, while analyses of multi-agent failures identify inter-agent misalignment and verification or termination problems~\citep{metagpt2023,mast2025}. In \rac, scoped contracts define the selected agent's task, and artifact-grounded verification informs subsequent work without vetoing transitions. Because additional coordination can increase inference costs, we follow cost-aware agent evaluation principles~\citep{agentsthatmatter2024,hal2025} and compare research scores, token usage, and estimated costs under matched run-level budgets.

\section{Runtime Agent Coordination for AI Scientists}

\subsection{Problem Formulation}

At decision step $t$, an AI scientist has a research state $s_t$ comprising its current objective, saved artifacts, unresolved problems, execution history, and remaining budget. Let $\mathcal{A}_t$ denote the agents currently permitted by the host. Communication adds information to this state, while runtime coordination uses it to select the next agent and specify its task. At a nonterminal handoff, the next agent is selected as
\begin{equation}
\label{eq:next_agent_selection}
a_{t+1} =
\begin{cases}
\pi_{\mathrm{host}}(s_t), & \text{native control (N0, R1)},\\[3pt]
\pi_{\mathrm{RAC}}(s_t,\mathcal{A}_t), & \text{runtime coordination (R2, R3)},
\end{cases}
\qquad a_{t+1}\in\mathcal{A}_t .
\end{equation}
Here, $\pi_{\mathrm{host}}$ is the host's native successor policy, and $\pi_{\mathrm{RAC}}$ is the shared runtime selection policy. Native policies may already include conditional branches, fallback rules, or adaptive delegation. Our comparison examines the effect of replacing this native selection policy with a shared state-dependent policy while retaining the host's permitted agents. Each condition generates its own research trajectory, so $s_t$ need not be identical across conditions.

\subsection{Engineering Implementation}

\begin{figure}[H]
\centering
\includegraphics[width=\textwidth]{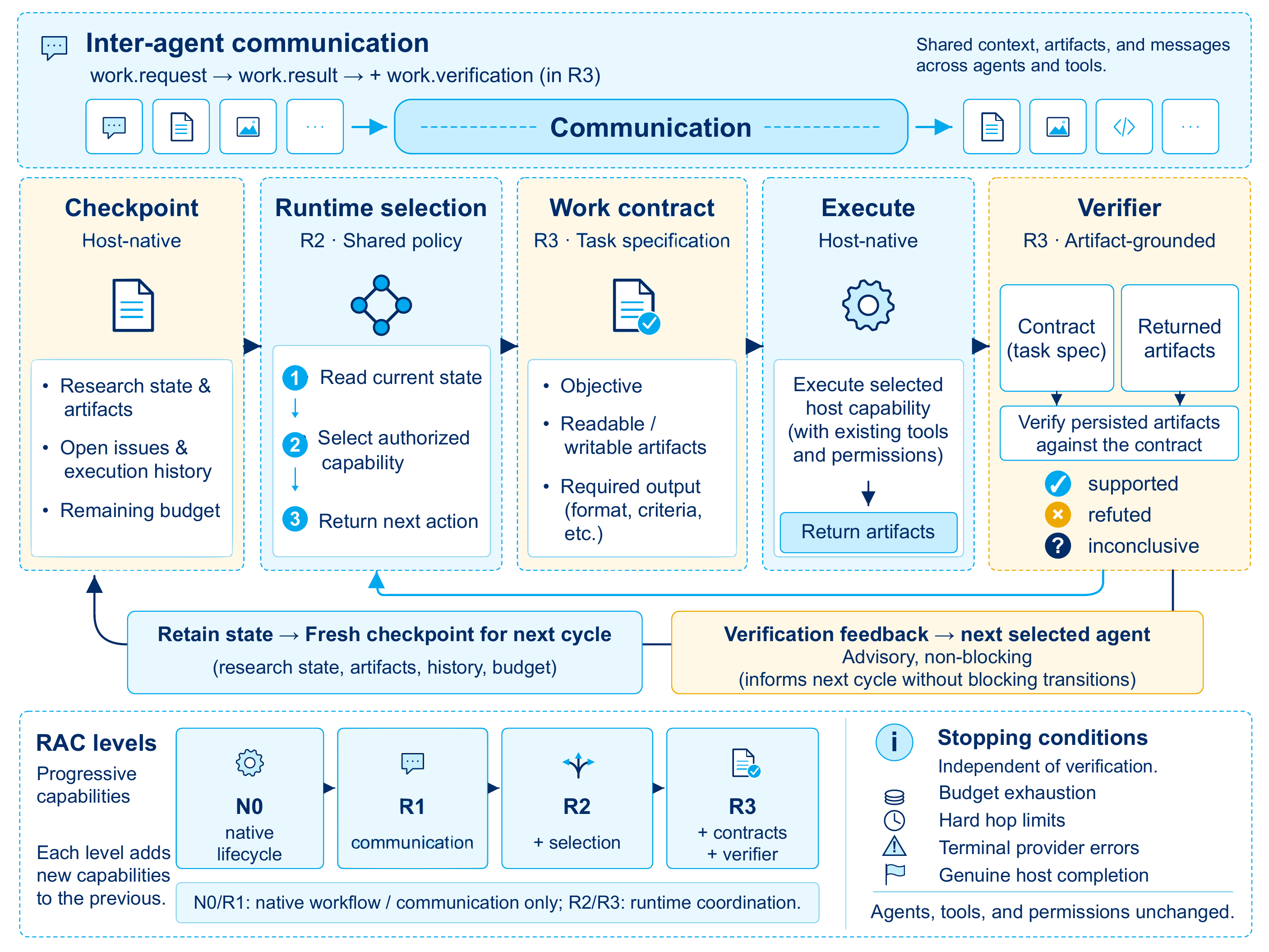}
\caption{\textbf{The RAC execution cycle.} At each handoff, agents exchange requests and results. In R2--R3, the current agent selects the next authorized host capability from a fresh checkpoint, without a separate orchestrator. R3 additionally specifies a work contract and verifies the returned artifacts, passing feedback to the next selected agent.}
\label{fig:rac_method}
\end{figure}

Figure~\ref{fig:rac_method} summarizes the execution cycle. We use the same \rac package and condition configuration across Agent Laboratory, EvoScientist, and ARK. Each host bridge serializes native state, invokes an existing capability, and returns artifacts and usage records through a shared interface. Bridges contain no host-specific runtime selection rules, verification criteria, or benchmark hints. This design preserves each host's agents, tools, and permissions without requiring identical host internals.

N0 executes the host-native lifecycle without additional communication mechanisms or a \rac phase loop. R1 introduces inter-agent communication while preserving the native successor at every handoff. R2 retains this communication mechanism and enables the current agent to select the next authorized capability using artifacts, execution history, unresolved problems, and remaining budget. Agents exchange structured requests and results through a common communication channel. Thus, R1--R2 changes next-agent selection while retaining the same communication mechanism. Communication state is isolated between runs, and cross-run memory is reset across all conditions.

R3 adds scoped work contracts and artifact-grounded verification. Each contract specifies the objective, readable and writable artifacts, and required output. The verifier checks persisted artifacts against the contract and records a \texttt{supported}, \texttt{refuted}, or \texttt{inconclusive} verdict. All verdicts preserve the workspace and host transition; feedback enters the next selected agent's prompt rather than directly triggering retries, rerouting, rollback, or stopping. Runs terminate on budget exhaustion, hard hop limits, terminal provider errors, or native host completion. Checkpoints and event records document coordination decisions, while external benchmark evaluation assesses final outputs separately from in-loop verification.

\section{Experimental Setup}

\subsection{Baselines}

We evaluated \rac with Agent Laboratory~\citep{agentlab2025}, EvoScientist~\citep{evoscientist2026}, and ARK~\citep{xin2026ark}. These hosts span predefined control loops, fixed research phases, and native adaptation, allowing us to examine runtime coordination across different native control mechanisms. Each host's native lifecycle serves as its N0 baseline.

\subsection{Benchmark and Conditions}

ResearchClawBench provides real-paper-derived tasks across scientific domains. Each task supplies a research question, related literature, raw data, and an executable environment while withholding the target paper~\citep{researchclawbench2026}. Our evaluation covers mathematics, neuroscience, information science, energy, life science, and physics. We evaluate Agent Laboratory and EvoScientist on ten tasks each and ARK on five tasks. All runs use seed~0.

Table~\ref{tab:rac_ladder} defines the cumulative conditions used to examine communication, runtime selection, and contracts with verification. R3 versus N0 compares the full framework with native execution; R2--R3 treats contracts and verification as a joint addition rather than isolating their individual effects.

\begin{table}[H]
\centering
\caption{Four cumulative conditions. R1 preserves native successors, R2 adds shared runtime selection, and R3 jointly adds scoped contracts and verification.}
\label{tab:rac_ladder}
\small
\begin{tabularx}{\linewidth}{@{}p{0.7cm}p{3.1cm}>{\raggedright\arraybackslash}X@{}}
\toprule
ID & Configuration & Newly enabled mechanism \\
\midrule
N0 & Native lifecycle & Native execution without additional communication mechanisms or a \rac phase loop. \\
R1 & Communication & Inter-agent communication with the native successor at each handoff. \\
R2 & + Runtime selection & Select a host capability from a fresh checkpoint using the shared policy. \\
R3 & + Contracts and\newline verification & Scope the task and check artifacts; pass supported, refuted, or inconclusive feedback to the next selected agent. Retain state and transitions for all verdicts. \\
\bottomrule
\end{tabularx}
\end{table}

\subsection{Controls and Analysis}

Within each host, we hold base models, native role prompts, tools, permissions, artifact interfaces, and task inputs fixed across conditions. Hop ceilings are calibrated to the corresponding host's N0 hop count rather than set uniformly across hosts. The budget-matching design concerns configured resource ceilings, not equal realized consumption; routing and verification are charged to the run's budget. Budget accounting is described in Appendix~\ref{app:budget_accounting}. Cross-run memory and evolution stores are reset before each run to isolate within-run coordination. Model choices and inference-cost accounting are detailed in Appendix~\ref{sec:cost_accounting}.

The primary outcome is the official ResearchClawBench weighted final score. We report input and output token usage and estimated inference costs separately. Each task--host--condition contributes one final score from the consolidated results record, including numeric zeros. A zero is an evaluator outcome, not an encoding of a missing report or a failed evaluation. Within-task differences from N0 are averaged over the same evaluated tasks within each host. Table~\ref{tab:task_scores} provides task-level scores; all summaries use unrounded values before display rounding.

\section{Results}

\subsection{Full runtime coordination across host systems}

Table~\ref{tab:main_results} compares full \rac (R3) with native execution. R3 increased mean scores for Agent Laboratory and ARK, while EvoScientist remained nearly unchanged. Recorded input-token usage and estimated costs decreased for Agent Laboratory but increased for the other two hosts.

\begin{table}[H]
\centering
\caption{Means over recorded results, with equal task weighting of final evaluation scores. Mean total input tokens are shown in millions (M). Costs are standardized estimates in USD from recorded input and output tokens using the cost-accounting convention in Appendix~\ref{sec:cost_accounting}.}
\label{tab:main_results}
\small
\begin{tabular}{@{}llrrr@{}}
\toprule
Host & Condition & Mean score & \shortstack{Mean total\\input tokens (M)} & Est. cost (USD) \\
\midrule
ARK & Native (N0) & 16.66 & 48.85 & 69.31 \\
ARK & Full \rac (R3) & 17.98 & 68.76 & 96.69 \\
Agent Laboratory & Native (N0) & 5.47 & 4.26 & 7.09 \\
Agent Laboratory & Full \rac (R3) & 9.88 & 2.56 & 4.48 \\
EvoScientist & Native (N0) & 15.99 & 11.49 & 15.44 \\
EvoScientist & Full \rac (R3) & 15.96 & 38.58 & 51.50 \\
\bottomrule
\end{tabular}
\end{table}

\subsection{Cumulative condition comparison}

R2 had the highest observed mean for all three hosts (Table~\ref{tab:cross_host_placeholder}). R1 also exceeded N0 for Agent Laboratory and ARK but fell below it for EvoScientist. Adding contracts and verification in R3 reduced the mean relative to R2 in every host.

\begin{table}[H]
\centering
\caption{Mean scores across all available tasks for each host and condition. Tasks are equally weighted using their final evaluation scores; recorded zeros are retained.}
\label{tab:cross_host_placeholder}
\small
\begin{tabular}{@{}p{6.2cm}ccc@{}}
\toprule
Shared configuration & ARK & Agent Lab. & EvoScientist \\
\midrule
N0: Native lifecycle & 16.66 & 5.47 & 15.99 \\
R1: Runtime communication & 17.40 & 9.63 & 12.07 \\
R2: + Runtime selection & 18.42 & 12.08 & 18.53 \\
R3: + Contracts and verification & 17.98 & 9.88 & 15.96 \\
\bottomrule
\end{tabular}
\end{table}

Table~\ref{tab:task_scores} shows that improvements over N0 outnumber declines under R2 and R3, with the broadest gains under R2; R1 produces a more evenly mixed pattern. R3 improved \mbox{Math-000} and \mbox{Math-001} for both Agent Laboratory and EvoScientist, but reduced EvoScientist's scores on \mbox{Physics-002} and \mbox{Math-002} and \mbox{Math-003}. Gains persisted across all three conditions for ARK on \mbox{Physics-002} and Agent Laboratory on \mbox{Life-001}.

\begin{table}
\centering
\caption{Task-level final scores at seed~0. The upper block reports the five tasks evaluated with all three hosts. The lower block reports five additional tasks evaluated with Agent Laboratory and EvoScientist. Each cell gives the score; R1--R3 additionally show the change from the same host's N0 in brackets. Dark green and red denote gains and declines exceeding 3 points, respectively; smaller changes remain uncolored. Colors describe magnitude, not statistical significance.}
\label{tab:task_scores}
\small
\setlength{\tabcolsep}{2pt}
\renewcommand{\arraystretch}{1.05}
\begin{tabularx}{\linewidth}{@{}ll*{5}{>{\centering\arraybackslash}X}@{}}
\toprule
Host & Condition & \shortstack{Neuroscience\\ 000} & \shortstack{Energy\\ 001} & \shortstack{Life\\ 001} & \shortstack{Neuroscience\\ 002} & \shortstack{Physics\\ 002} \\
\midrule
\multirow{4}{*}{ARK} & N0 & 10.00 & 29.30 & 9.80 & 6.60 & 27.60 \\
 & R1 & \taskscore{7.60}{-2.40} & \taskscore{27.60}{-1.70} & \taskscore{9.40}{-0.40} & \taskscore{5.70}{-0.90} & \taskscore{36.70}{+9.10} \\
 & R2 & \taskscore{8.40}{-1.60} & \taskscore{31.90}{+2.60} & \taskscore{14.10}{+4.30} & \taskscore{5.70}{-0.90} & \taskscore{32.00}{+4.40} \\
 & R3 & \taskscore{13.80}{+3.80} & \taskscore{27.90}{-1.40} & \taskscore{10.45}{+0.65} & \taskscore{3.15}{-3.45} & \taskscore{34.60}{+7.00} \\
\midrule
\multirow{4}{*}{Agent Lab.} & N0 & 9.00 & 9.20 & 1.30 & 3.30 & 0.00 \\
 & R1 & \taskscore{8.00}{-1.00} & \taskscore{10.50}{+1.30} & \taskscore{4.65}{+3.35} & \taskscore{5.40}{+2.10} & \taskscore{6.60}{+6.60} \\
 & R2 & \taskscore{3.40}{-5.60} & \taskscore{12.20}{+3.00} & \taskscore{6.90}{+5.60} & \taskscore{5.70}{+2.40} & \taskscore{24.55}{+24.55} \\
 & R3 & \taskscore{10.40}{+1.40} & \taskscore{5.90}{-3.30} & \taskscore{2.90}{+1.60} & \taskscore{5.05}{+1.75} & \taskscore{4.30}{+4.30} \\
\midrule
\multirow{4}{*}{EvoScientist} & N0 & 8.40 & 17.70 & 16.20 & 7.80 & 39.20 \\
 & R1 & \taskscore{6.60}{-1.80} & \taskscore{19.90}{+2.20} & \taskscore{13.50}{-2.70} & \taskscore{4.60}{-3.20} & \taskscore{10.40}{-28.80} \\
 & R2 & \taskscore{6.00}{-2.40} & \taskscore{25.80}{+8.10} & \taskscore{13.75}{-2.45} & \taskscore{8.30}{+0.50} & \taskscore{43.10}{+3.90} \\
 & R3 & \taskscore{13.60}{+5.20} & \taskscore{28.50}{+10.80} & \taskscore{14.10}{-2.10} & \taskscore{7.05}{-0.75} & \taskscore{16.45}{-22.75} \\
\bottomrule
\end{tabularx}

\vspace{0.8em}
\begin{tabularx}{\linewidth}{@{}ll*{5}{>{\centering\arraybackslash}X}@{}}
\toprule
Host & Condition & \shortstack{Math\\ 000} & \shortstack{Math\\ 001} & \shortstack{Math\\ 002} & \shortstack{Math\\ 003} & \shortstack{Information\\ 003} \\
\midrule
\multirow{4}{*}{Agent Lab.} & N0 & 17.15 & 11.10 & 0.00 & 0.00 & 3.60 \\
 & R1 & \taskscore{18.90}{+1.75} & \taskscore{23.10}{+12.00} & \taskscore{3.60}{+3.60} & \taskscore{10.05}{+10.05} & \taskscore{5.45}{+1.85} \\
 & R2 & \taskscore{22.50}{+5.35} & \taskscore{27.50}{+16.40} & \taskscore{8.40}{+8.40} & \taskscore{6.60}{+6.60} & \taskscore{3.00}{-0.60} \\
 & R3 & \taskscore{18.30}{+1.15} & \taskscore{29.90}{+18.80} & \taskscore{4.80}{+4.80} & \taskscore{9.00}{+9.00} & \taskscore{8.20}{+4.60} \\
\midrule
\multirow{4}{*}{EvoScientist} & N0 & 16.05 & 16.80 & 10.60 & 16.30 & 10.80 \\
 & R1 & \taskscore{9.60}{-6.45} & \taskscore{32.60}{+15.80} & \taskscore{5.30}{-5.30} & \taskscore{9.45}{-6.85} & \taskscore{8.70}{-2.10} \\
 & R2 & \taskscore{21.60}{+5.55} & \taskscore{36.20}{+19.40} & \taskscore{11.00}{+0.40} & \taskscore{11.25}{-5.05} & \taskscore{8.30}{-2.50} \\
 & R3 & \taskscore{22.35}{+6.30} & \taskscore{37.70}{+20.90} & \taskscore{3.20}{-7.40} & \taskscore{9.70}{-6.60} & \taskscore{6.90}{-3.90} \\
\bottomrule
\end{tabularx}
\end{table}

\begin{figure}
\centering
\includegraphics[width=\textwidth]{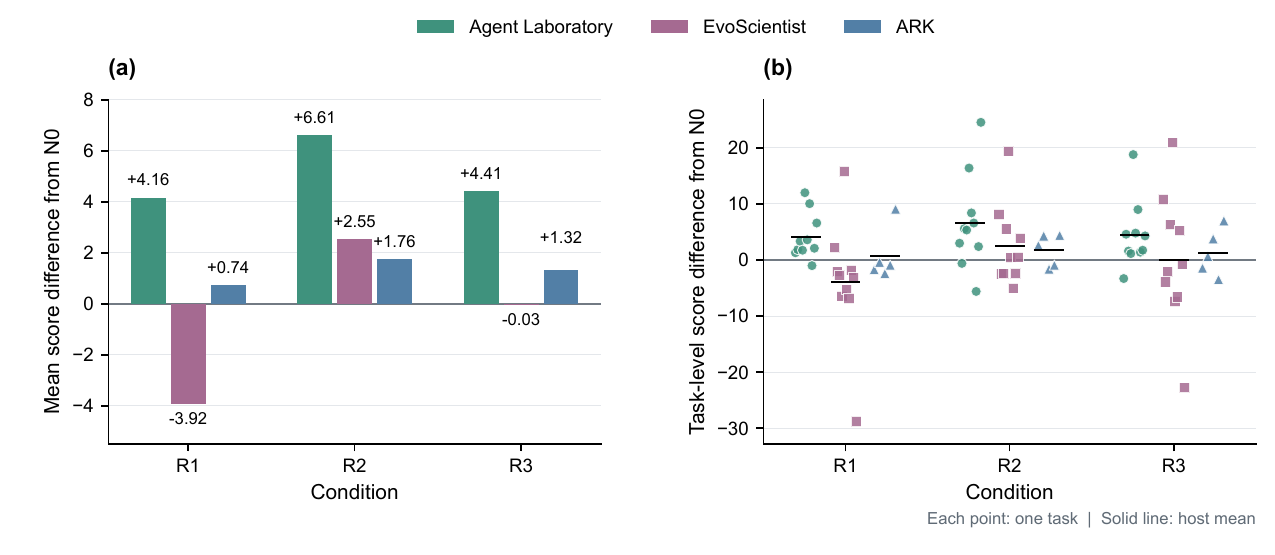}
\caption{\textbf{Within-task score improvements.} (a) Mean score differences from N0 for each host under R1--R3. (b) Individual task-level differences from the same host's N0; solid black segments indicate host-specific means. Points are horizontally offset for visibility. All conditions use the same task set within each host. The panels use different vertical scales and show descriptive results, not uncertainty across independent runs.}
\label{fig:score_conditions}
\end{figure}

Figure~\ref{fig:score_conditions} contrasts mean improvements with their task-level spread. R2 increased scores over N0 by 6.61 points for Agent Laboratory, 2.55 for EvoScientist, and 1.76 for ARK. Both gains and declines occurred within every host and condition, showing that positive means did not imply uniform benefits.

Detailed statistics of token usage and estimated inference costs are reported in Appendix~\ref{app:token_usage}.

\subsection{Transferability Test}

\begin{figure}[H]
\centering
\includegraphics[width=\linewidth]{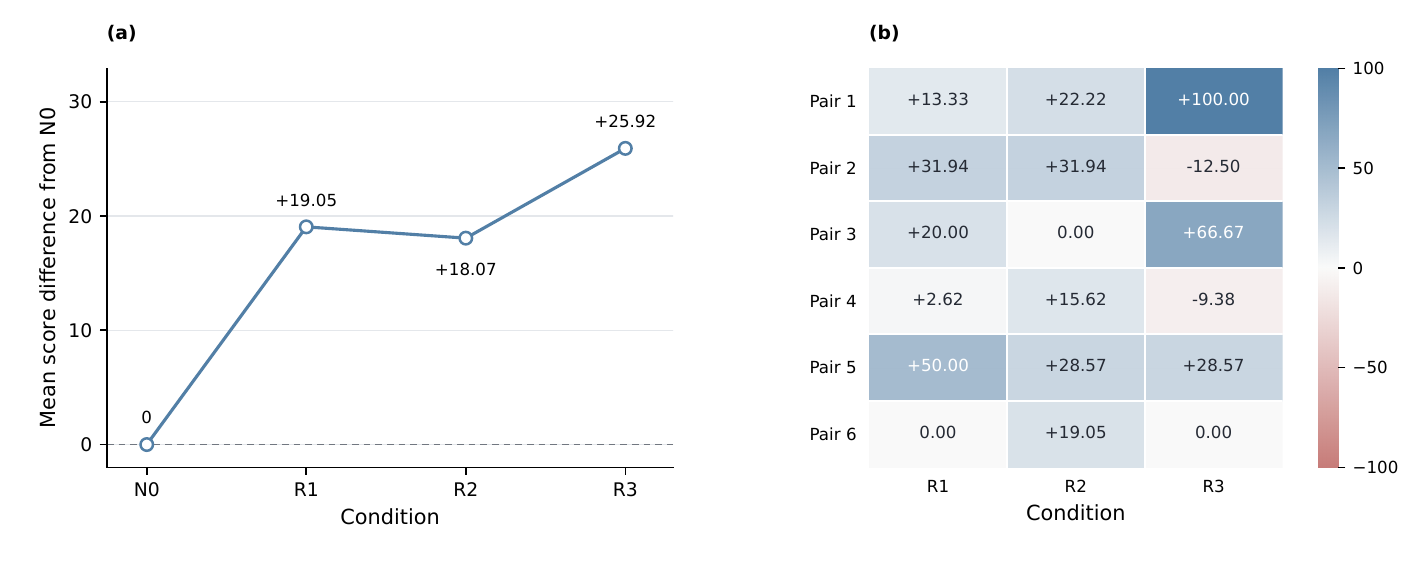}
\caption{\textbf{Transfer to DiscoveryBench.}
(a) Mean score improvements over N0, computed within each host and task, averaged across available hosts within each task, and then equally across the five tasks. (b) Improvements for the six task--host pairs, labeled Pairs 1--6; blue indicates gains and red indicates decreases. Table~\ref{tab:discoverybench_scores} lists the tasks, hosts, and absolute scores; the first two rows share the same task. Scores are rescaled to 0--100; the improvement relative to N0 is zero by construction for N0. Points in (a) show descriptive means, and connecting lines are visual guides.}
\label{fig:discoverybench_transfer}
\end{figure}

We additionally evaluated \rac on an exploratory subset of five DiscoveryBench~\citep{discoverybench2025} tasks, without a predefined sampling rule. Agent Laboratory covers all five tasks; EvoScientist covers only NLS SES, giving six task--host pairs rather than a complete two-host matrix. Figure~\ref{fig:discoverybench_transfer} shows mean improvements over N0 of 19.05, 18.07, and 25.92 points under R1--R3, respectively. R3 achieved the largest aggregate gain, extending the evaluation beyond ResearchClawBench without establishing representative cross-benchmark performance.

R2's slightly lower mean than R1 was driven by the incarceration and archaeology tasks, despite improvements on the other three tasks. In incarceration, R2 analyzed simulated data and reported synthetic year indices rather than the requested survey years. In archaeology, additional quantities in R2's answer reduced the evaluator's variable-matching score, while relation and context matching remained unchanged. Both runs completed the same six-stage sequence as R1 without exhausting their budgets. These cases indicate task drift and scoring sensitivity rather than a demonstrated disadvantage of runtime selection.

\subsection{Trace-level analysis: three coordination effects}
\label{sec:trace_effects}

We inspected 1,181 hop excerpts and 180 external scoring receipts from the 60 cells covering five shared tasks, three hosts, and N0--R3. Selected trajectories illustrate coordination mechanisms and failure modes, not their prevalence or the causes of aggregate score differences. The excerpts are incomplete, scientific claims were not independently recomputed, and hop indices are zero-based.

\paragraph{Evidence--action closure effect.}

In \mbox{Physics-002} with EvoScientist R2, the native reviewer reported at hop~6 that the report contained only 39 bytes despite existing code, analyses, and figures. At hop~7, the selected debug capability reported producing a 337-line report of approximately 22\,KB. Ten recorded scientific-artifact changes and the final inventory support the conclusion that a report was produced. This handoff connected a specific defect to corrective work, although subsequent debug calls made no recorded scientific-artifact changes before the hop limit. The feedback came from the host reviewer, not R3's verifier; delivery repair does not establish scientific correctness.

\paragraph{Planning stagnation after diagnosis.}
In \mbox{Physics-002} with EvoScientist R3, the reviewer identified a standard-error discrepancy: 0.011 in the report versus 0.015 in the data. At hop~8, the planner attributed it to duplicated observations and proposed code changes followed by regeneration of results. However, hops~8--13 remained planner calls with no recorded scientific-artifact changes, ending at the 14-hop limit. The trajectory illustrates a failure to move from diagnosis to implementation. Claims of repeated verification appeared in agent messages but were not independently confirmed from verifier events.

\paragraph{Self-consistent drift effect.}
In \mbox{Neuroscience-000} with EvoScientist R3, the native reviewer reported that numerical and code--report consistency checks passed and assigned a score of 8/10. External evaluation nevertheless identified missing cross-laboratory, sex, and environment SHAP comparisons, yielding a final score of 13.6. The two scores use different criteria and scales and are not directly comparable. Similarly, \mbox{Life-001} with Agent Laboratory R2 analyzed ten synthetic patients rather than the supplied seven-patient dataset. These cases distinguish internal consistency from coverage of the research objective, without identifying whether divergence originated in task interpretation, data limitations, or coordination.

Appendix~\ref{app:rac_trace_cases} provides detailed trajectories and evidence boundaries, including additional Math cases outside the shared-task subset.

\section{Discussion}

Runtime selection achieved the highest observed means on ResearchClawBench, whereas contracts and verification produced the largest aggregate gain on the exploratory DiscoveryBench subset. This contrast suggests that the benefits of additional coordination depend on the host and research setting, rather than increasing uniformly with each mechanism.

Under matched budgets, communication and verification add context and inference costs that may displace useful research work. Runtime selection can offset this overhead by directing work toward relevant agents, but R2's lower recorded costs in ARK and Agent Laboratory did not extend to EvoScientist. Additional budget might accommodate R3's overhead, yet the repeated planning trajectory in Section~\ref{sec:trace_effects} shows that diagnosis need not lead to implementation. Distracting context or inaccurate verifier feedback could also weaken outcomes. Budget sweeps with per-hop usage and artifact changes would help distinguish resource constraints from ineffective delegation.

Host structure may further shape these trade-offs. Agent Laboratory's fixed phases may benefit from information carried between stages, whereas EvoScientist already incorporates native adaptation. Differences in native prompts and token consumption may also affect how effectively hosts use additional context. These remain hypotheses: aggregate scores and usage do not establish why individual runs improve or decline.

Our single-seed evaluation does not measure variability across independent runs, and bounded benchmark tasks do not establish performance in open-ended discovery. Costs are standardized token-based estimates rather than provider invoices. R3 jointly adds contracts and verification, so their individual effects remain unresolved.

\section{Conclusion}

We presented \rac as an alternative coordination approach for multi-agent AI scientists. It selects existing agents from current state and, in R3, adds scoped contracts and artifact-grounded verification feedback. Across all available ResearchClawBench scores, R2 had the highest task-weighted mean for each host, whereas R3's performance relative to N0 varied across hosts. These rankings are descriptive and do not establish robustness across independent runs. Standardized token-based costs describe resource use without establishing a general performance or cost advantage.

\subsection*{Ethics and reproducibility statement}
Adaptive scientific workflows may accelerate research but can also amplify unsupported claims or optimize for fallible reviewer signals. \rac preserves host permissions, scopes tasks through contracts, and records verification evidence, but verification is not a safety gate and does not prevent unsupported outputs from propagating; accountable human review remains necessary, particularly in safety-critical, dual-use, or regulated domains. Sections~3 and~4 describe the host adaptation boundary and experimental conditions, Table~\ref{tab:task_scores} reports task-level scores, Appendix~\ref{sec:cost_accounting} specifies the execution and judge models and token-based cost calculation, and Appendix~\ref{app:analysis_spec} defines task weighting and the paired comparisons in Figure~\ref{fig:score_conditions}. The implementation is available at \url{https://github.com/systemind-team/Runtime-AI-Scientist}.

\clearpage
\bibliography{rac_ai_scientist}

@article{aiscientist2024,
  title={{The AI Scientist: Towards Fully Automated Open-Ended Scientific Discovery}},
  author={Chris Lu and Cong Lu and Robert Tjarko Lange and Jakob Foerster and Jeff Clune and David Ha},
  journal={arXiv preprint arXiv:2408.06292}, year={2024},
  url={https://arxiv.org/abs/2408.06292}
}

@article{aiscientistv2,
  title={{The AI Scientist-v2: Workshop-Level Automated Scientific Discovery via Agentic Tree Search}},
  author={Yutaro Yamada and Robert Tjarko Lange and Cong Lu and Shengran Hu and Chris Lu and Jakob Foerster and Jeff Clune and David Ha},
  journal={arXiv preprint arXiv:2504.08066}, year={2025},
  url={https://arxiv.org/abs/2504.08066}
}

@article{agentlab2025,
  title={{Agent Laboratory: Using LLM Agents as Research Assistants}},
  author={Samuel Schmidgall and Yusheng Su and Ze Wang and Ximeng Sun and Jialian Wu and Xiaodong Yu and others},
  journal={arXiv preprint arXiv:2501.04227}, year={2025},
  url={https://arxiv.org/abs/2501.04227}
}

@article{coscientist2025,
  title={{Accelerating Scientific Discovery with Co-Scientist}},
  author={Juraj Gottweis and Wei-Hung Weng and Alexander Daryin and Tao Tu and Petar Sirkovic and Artiom Myaskovsky and others},
  journal={arXiv preprint arXiv:2502.18864}, year={2025},
  url={https://arxiv.org/abs/2502.18864}
}

@article{aris2026,
  title={{ARIS: Autonomous Research via Adversarial Multi-Agent Collaboration}},
  author={Ruofeng Yang and Yongcan Li and Shuai Li},
  journal={arXiv preprint arXiv:2605.03042}, year={2026},
  url={https://arxiv.org/abs/2605.03042}
}

@article{autoscientists2026,
  title={{AutoScientists: Self-Organizing Agent Teams for Long-Running Scientific Experimentation}},
  author={Shanghua Gao and Ada Fang and Marinka Zitnik},
  journal={arXiv preprint arXiv:2605.28655}, year={2026},
  url={https://arxiv.org/abs/2605.28655}
}

@article{freephdlabor2025,
  title={{Build Your Personalized Research Group: A Multiagent Framework for Continual and Interactive Science Automation}},
  author={Ed Li and Junyu Ren and Xintian Pan and Cat Yan and Chuanhao Li and Dirk Bergemann and Zhuoran Yang},
  journal={arXiv preprint arXiv:2510.15624}, year={2025},
  url={https://arxiv.org/abs/2510.15624}
}

@inproceedings{gptswarm2024,
  author={Mingchen Zhuge and Wenyi Wang and Louis Kirsch and Francesco Faccio and Dmitrii Khizbullin and J{\"u}rgen Schmidhuber},
  title={{GPTSwarm}: Language Agents as Optimizable Graphs},
  booktitle={Proceedings of the 41st International Conference on Machine Learning},
  volume={235}, pages={62743--62767}, publisher={PMLR}, year={2024},
  url={https://proceedings.mlr.press/v235/zhuge24a.html}
}

@article{dylan2023,
  author={Zijun Liu and Yanzhe Zhang and Peng Li and Yang Liu and Diyi Yang},
  title={A Dynamic {LLM}-Powered Agent Network for Task-Oriented Agent Collaboration},
  journal={arXiv preprint arXiv:2310.02170}, year={2023},
  url={https://arxiv.org/abs/2310.02170}
}

@article{researchclawbench2026,
  title={{ResearchClawBench: A Benchmark for End-to-End Autonomous Scientific Research}},
  author={Wanghan Xu and Shuo Li and Tianlin Ye and Qinglong Cao and Yixin Chen and Hengjian Gao and others},
  journal={arXiv preprint arXiv:2606.07591}, year={2026},
  url={https://arxiv.org/abs/2606.07591}
}

@article{autoresearchfail2026,
  title={{How Do Agents Fail on AutoResearch: End-to-End Diagnostic Evaluation on 100 Real-World Frontier Research Tasks}},
  author={Yanlin Fei and Nazhou Liu and Xinmiao Yu and Shaolong Chen and Lei Li and Rahul Thapa and others},
  journal={arXiv preprint arXiv:2608.14905}, year={2026},
  url={https://arxiv.org/abs/2608.14905}
}

@article{agentsthatmatter2024,
  title={{AI Agents That Matter}},
  author={Sayash Kapoor and Benedikt Stroebl and Zachary S. Siegel and Nitya Nadgir and Arvind Narayanan},
  journal={arXiv preprint arXiv:2407.01502}, year={2024},
  url={https://arxiv.org/abs/2407.01502}
}

@misc{hal2025,
  title={{HAL: Holistic Agent Leaderboard}},
  author={Sayash Kapoor and Benedikt Stroebl and Peter Kirgis and Nitya Nadgir and Zachary S. Siegel and others},
  howpublished={\url{https://hal.cs.princeton.edu}}, year={2025}
}

@article{mast2025,
  author={Mert Cemri and Melissa Z. Pan and Shuyi Yang and Lakshya A. Agrawal and Bhavya Chopra and Rishabh Tiwari and others},
  title={Why Do Multi-Agent {LLM} Systems Fail?},
  journal={arXiv preprint arXiv:2503.13657}, year={2025},
  url={https://arxiv.org/abs/2503.13657}
}

@article{canaievaluate2026,
  title={{Can AI Evaluate AI Scientists? A Benchmarking Study of Autonomous Research Generation Systems Using Automated Multi-Model Review}},
  author={Vaibhava Lakshmi Ravideshik and Mayank Kejriwal},
  journal={arXiv preprint arXiv:2607.28631}, year={2026},
  url={https://arxiv.org/abs/2607.28631}
}

@inproceedings{xin2026ark,
  title={{ARK}: A Human-Steered Research Harness for {AI}-Assisted Scientific Discovery},
  author={Jihao Xin and Yangzhixin Luo and Muhammad Bilal and Marco Canini},
  booktitle={ICML 2026 AI for Science Workshop}, year={2026},
  url={https://openreview.net/forum?id=fSAiWmpk0O}
}

@article{airesearcher2025,
  title={{AI-Researcher}: Autonomous Scientific Innovation},
  author={Tang, Jiabin and Xia, Lianghao and Li, Zhonghang and Huang, Chao},
  journal={arXiv preprint arXiv:2505.18705},
  year={2025},
  url={https://arxiv.org/abs/2505.18705}
}

@article{autoresearchclaw2026,
  title={{AutoResearchClaw}: Self-Reinforcing Autonomous Research with Human-{AI} Collaboration},
  author={Liu, Jiaqi and Qiu, Shi and Li, Mairui and Li, Bingzhou and Ji, Haonian and Han, Siwei and Ye, Xinyu and Xia, Peng and Dong, Zihan and Chen, Meng and Zhang, Congyu and others},
  journal={arXiv preprint arXiv:2605.20025},
  year={2026},
  url={https://arxiv.org/abs/2605.20025}
}

@article{goel2025training,
  title={Training {AI} co-scientists using rubric rewards},
  author={Goel, Shashwat and Hazra, Rishi and Jayalath, Dulhan and Willi, Timon and Jain, Parag and Shen, William F and Leontiadis, Ilias and Barbieri, Francesco and Bachrach, Yoram and Geiping, Jonas and others},
  journal={arXiv preprint arXiv:2512.23707},
  year={2025},
  url={https://arxiv.org/abs/2512.23707}
}

@article{autogen2023,
  title={{AutoGen}: Enabling Next-Gen {LLM} Applications via Multi-Agent Conversation},
  author={Wu, Qingyun and Bansal, Gagan and Zhang, Jieyu and Wu, Yiran and Li, Beibin and Zhu, Erkang and Jiang, Li and Zhang, Xiaoyun and Zhang, Shaokun and Liu, Jiale and Awadallah, Ahmed Hassan and White, Ryen W. and Burger, Doug and Wang, Chi},
  journal={arXiv preprint arXiv:2308.08155},
  year={2023},
  url={https://arxiv.org/abs/2308.08155}
}

@article{metagpt2023,
  title={{MetaGPT}: Meta Programming for a Multi-Agent Collaborative Framework},
  author={Hong, Sirui and Zhuge, Mingchen and Chen, Jiaqi and Zheng, Xiawu and Cheng, Yuheng and Zhang, Ceyao and Wang, Jinlin and Wang, Zili and Yau, Steven Ka Shing and Lin, Zijuan and Zhou, Liyang and Ran, Chenyu and Xiao, Lingfeng and Wu, Chenglin and Schmidhuber, J{\"u}rgen},
  journal={arXiv preprint arXiv:2308.00352},
  year={2023},
  url={https://arxiv.org/abs/2308.00352}
}

@article{du2023debate,
  title={Improving Factuality and Reasoning in Language Models through Multiagent Debate},
  author={Du, Yilun and Li, Shuang and Torralba, Antonio and Tenenbaum, Joshua B. and Mordatch, Igor},
  journal={arXiv preprint arXiv:2305.14325},
  year={2023},
  url={https://arxiv.org/abs/2305.14325}
}

@inproceedings{discoverybench2025,
  title={{DiscoveryBench}: Towards Data-Driven Discovery with Large Language Models},
  author={Majumder, Bodhisattwa Prasad and Surana, Harshit and Agarwal, Dhruv and Mishra, Bhavana Dalvi and Meena, Abhijeetsingh and Prakhar, Aryan and Vora, Tirth and Khot, Tushar and Sabharwal, Ashish and Clark, Peter},
  booktitle={International Conference on Learning Representations},
  year={2025},
  url={https://proceedings.iclr.cc/paper_files/paper/2025/hash/0d70af566e69f1dfb687791ecf955e28-Abstract-Conference.html}
}

@article{evoscientist2026,
  title={{EvoScientist}: Towards Multi-Agent Evolving {AI} Scientists for End-to-End Scientific Discovery},
  author={Lyu, Yougang and Zhang, Xi and Yi, Xinhao and Zhao, Yuyue and Guo, Shuyu and Hu, Wenxiang and Piotrowski, Jan and Kaliski, Jakub and Urbani, Jacopo and Meng, Zaiqiao and Zhou, Lun and Yan, Xiaohui},
  journal={arXiv preprint arXiv:2603.08127},
  year={2026},
  url={https://arxiv.org/abs/2603.08127}
}

@inproceedings{wang2026hardcoding,
  title={Position: Stop Hardcoding Multi-Agent Workflows That General Agents Will Outgrow},
  author={Xisen Wang},
  booktitle={ICML 2026 Workshop on Agents in the Wild: Safety, Security, and Beyond},
  year={2026},
  url={https://openreview.net/forum?id=Icz1RJ2t0j}
}
\bibliographystyle{systemind_research}

\clearpage
\appendix

\newcommand{\appcontentssec}[2]{\par\vspace{5pt}\noindent\hyperref[#1]{\textbf{\makebox[1.6em][l]{\ref*{#1}}#2}}\nobreak\dotfill\nobreak\textbf{\pageref*{#1}}\par}
\newcommand{\appcontentssub}[2]{\par\vspace{1pt}\noindent\hspace*{1.6em}\hyperref[#1]{\makebox[2.4em][l]{\ref*{#1}}#2}\nobreak\dotfill\nobreak\pageref*{#1}\par}

\section*{Appendix Contents}
\begingroup
\hypersetup{pdfborder={0 0 0}}
\small
\appcontentssec{app:implementation}{Implementation Details}
\appcontentssub{app:records}{Coordination records}
\appcontentssub{app:bridges}{Host bridges and termination}
\appcontentssec{sec:cost_accounting}{Experimental Details}
\appcontentssub{app:models}{Models}
\appcontentssub{app:budget_accounting}{Budget calibration and accounting}
\appcontentssub{app:cost_formula}{Cost accounting}
\appcontentssub{app:analysis_spec}{Evaluation protocol}
\appcontentssec{app:additional_results}{Additional Results}
\appcontentssub{app:token_usage}{Token usage and estimated cost}
\appcontentssub{app:discoverybench_scores}{DiscoveryBench task-level scores}
\appcontentssub{app:criterion_scores}{ResearchClawBench criterion-level scores}
\appcontentssub{app:criterion_summaries}{ResearchClawBench criterion summaries}
\appcontentssec{app:rac_trace_cases}{Detailed Case Studies of RAC Coordination}
\appcontentssub{app:case_ark_life}{ARK on \mbox{Life-001}: direct corrective handoffs}
\appcontentssub{app:case_ark_physics}{ARK on \mbox{Physics-002}: repair followed by cross-capability checking}
\appcontentssub{app:case_evo_math001}{EvoScientist on \mbox{Math-001}: explicit delivery verification}
\appcontentssub{app:case_agentlab_math003}{Agent Laboratory on \mbox{Math-003}: greater task relevance}
\appcontentssub{app:case_evo_math003}{EvoScientist on \mbox{Math-003}: local contracts, unfinished objective}
\appcontentssec{app:all_hops_compact}{Condensed All-Hop Trajectories}
\appcontentssub{app:compact_physics}{\mbox{Physics-002} / ARK / R3}
\appcontentssub{app:compact_life}{\mbox{Life-001} / ARK / R2}
\appcontentssec{app:genai}{Generative AI Use Disclosure}
\endgroup

\clearpage

\section{Implementation Details}
\label{app:implementation}

This appendix specifies the records that \rac writes during a run and the boundary between \rac and each host. All conditions share the same \rac package and condition configuration; only the host bridge differs across hosts.

\subsection{Coordination records}
\label{app:records}

Table~\ref{tab:records} lists the records that document coordination decisions. Requests and results are typed \texttt{work.request} and \texttt{work.result} events exchanged through the common communication channel. R3 adds contracts and \texttt{work.verification} records. Every verdict is retained and passed to the next selected agent; no verdict discards the workspace or alters the host transition.

\begin{table}[H]
\centering
\caption{Coordination records written by \rac. Checkpoints and event records document coordination decisions; external benchmark evaluation assesses final outputs separately.}
\label{tab:records}
\small
\begin{tabularx}{\linewidth}{@{}>{\raggedright\arraybackslash}p{3.3cm}>{\raggedright\arraybackslash}X>{\raggedright\arraybackslash}p{3.4cm}@{}}
\toprule
Record & Contents & Use \\
\midrule
Checkpoint & Run and hop identifiers, objective, native state, artifacts, unresolved problems, remaining budget, and available capabilities. & State from which R2--R3 select the next authorized capability. \\
\texttt{work.request} & Structured request to the selected agent. & Hands off work through the communication channel. \\
\texttt{work.result} & Returned result with artifacts and usage records. & Returns work and charges usage to the run. \\
Contract (R3) & Objective, readable and writable artifacts, and required output. & Scopes the selected agent's task. \\
\texttt{work.verification}\newline (R3) & Verdict (\texttt{supported}, \texttt{refuted}, or \texttt{inconclusive}), checks, reasons, and feedback status. & Feedback for the next selected agent's prompt. \\
Provenance & Communication backend and non-secret room identifier; never invite tokens. & Attributes each trace to its channel. \\
\bottomrule
\end{tabularx}
\end{table}

\subsection{Host bridges and termination}
\label{app:bridges}

Each host bridge serializes native state, invokes an existing capability, and returns artifacts and usage records through a shared interface. Bridges contain no host-specific runtime selection rules, verification criteria, or benchmark hints, so each host keeps its agents, tools, and permissions. Verification feedback enters the next selected agent's prompt rather than directly triggering retries, rerouting, rollback, or stopping. Runs terminate on budget exhaustion, hard hop limits, terminal provider errors, or native host completion. Communication state is isolated between runs, and cross-run memory and evolution stores are reset before each run.

\section{Experimental Details}
\label{sec:cost_accounting}

This appendix gives the models, budget calibration, cost accounting, and evaluation protocol used for all reported comparisons.

\subsection{Models}
\label{app:models}
We use DeepSeek V4 Pro as the execution model for the AI-scientist hosts and GPT-5.5 as the external ResearchClawBench judge. External judging is separate from the in-loop verifier and does not control agent selection or host transitions.

\subsection{Budget calibration and accounting}
\label{app:budget_accounting}
Hop ceilings are tailored to each host using its N0 hop count as the calibration reference. They are not a common cross-host constant: a native phase may itself contain many model or tool calls. Configured limits and realized usage must therefore be distinguished. Token and call accounting includes coordination and verification, while external judging is excluded. Limits checked after an agent returns can be exceeded by that completed call; the excess is retained in usage totals, and subsequent work is stopped.

The archived configurations contain multiple ceiling settings, and the consolidated score table alone does not identify every final configuration. We therefore do not infer a single numerical budget vector for an entire host or equate the standardized USD estimates with an enforced provider-spend limit. Budget matching describes the intended comparison design, not an independently verified equality of every archived resource ceiling. The results should be read as exploratory comparisons rather than exact compute-equated effects.

\subsection{Cost accounting}
\label{app:cost_formula}
We estimate inference costs from recorded input and output token usage using a common DeepSeek V4 Pro tariff. We apply the official peak rates of US\$1.32 per million non-cached input tokens and US\$3.96 per million output tokens, as listed in the \href{https://api-docs.deepseek.com/quick_start/pricing/}{DeepSeek API pricing documentation} on 23 September 2026. For each episode, the estimated cost is $(1.32\,T_{\mathrm{in}}+3.96\,T_{\mathrm{out}})/10^6$ USD. Here, $T_{\mathrm{in}}$ and $T_{\mathrm{out}}$ are its recorded input and output token counts. We apply no cache-hit or off-peak discounts. Table entries average episode estimates before rounding. These are standardized inference-cost estimates, not Azure invoices, and exclude external GPT-5.5 judging charges.

\subsection{Evaluation protocol}
\label{app:analysis_spec}
The primary comparison is R3 versus N0 under matched budgets. Condition means equally weight final task--host--condition scores. For each host and each of R1--R3, we subtract the N0 score on the same task and average these differences over the same evaluated task set within that host. Figure~\ref{fig:score_conditions} reports these descriptive paired means. The bars do not measure variability across independent runs. Adjacent conditions are compared on the same tasks, but R2--R3 does not separate contracts from verification.

\section{Additional Results}
\label{app:additional_results}

This appendix reports resource usage, task-level DiscoveryBench scores, and criterion-level ResearchClawBench scores that support the aggregates in the main text.

\subsection{Token usage and estimated cost}
\label{app:token_usage}

Table~\ref{tab:token_usage} summarizes mean input and output tokens and standardized inference costs for the evaluated tasks. Agent Laboratory has its lowest input-token and cost means in R2, but its lowest output-token mean in R1. EvoScientist has its lowest token and cost means in N0. ARK has its lowest input-token, output-token, and cost means in R2. These are within-host resource summaries; the hosts use different task sets.

\begin{table}[t]
\centering
\caption{Mean recorded token usage and estimated inference cost. Input tokens are in millions (M); output tokens are in thousands (K). $n_{\mathrm{usage}}/n_{\mathrm{score}}$ gives the number of task records contributing to resource and score means. Costs use the uniform tariff in Appendix~\ref{app:cost_formula} and exclude external judging. Hosts cover different tasks, and resource means for Agent Laboratory R3 and EvoScientist R1 have smaller denominators than their score means.}
\label{tab:token_usage}
\small
\begin{tabular}{@{}llrrrr@{}}
\toprule
Host & Condition & $n_{\mathrm{usage}}/n_{\mathrm{score}}$ & Input (M) & Output (K) & Est. cost (USD) \\
\midrule
Agent Laboratory & N0 & 10/10 & 4.26 & 369.42 & 7.09 \\
Agent Laboratory & R1 & 10/10 & 2.90 & 269.91 & 4.89 \\
Agent Laboratory & R2 & 10/10 & 2.36 & 308.18 & 4.33 \\
Agent Laboratory & R3 & 9/10 & 2.56 & 279.24 & 4.48 \\
\midrule
EvoScientist & N0 & 10/10 & 11.49 & 70.34 & 15.44 \\
EvoScientist & R1 & 8/10 & 18.31 & 126.37 & 24.67 \\
EvoScientist & R2 & 10/10 & 37.74 & 146.97 & 50.40 \\
EvoScientist & R3 & 10/10 & 38.58 & 145.30 & 51.50 \\
\midrule
ARK & N0 & 5/5 & 48.85 & 1,219.69 & 69.31 \\
ARK & R1 & 5/5 & 63.06 & 1,425.76 & 88.88 \\
ARK & R2 & 5/5 & 45.47 & 1,113.58 & 64.42 \\
ARK & R3 & 5/5 & 68.76 & 1,496.85 & 96.69 \\
\bottomrule
\end{tabular}
\end{table}

\subsection{DiscoveryBench task-level scores}
\label{app:discoverybench_scores}

Table~\ref{tab:discoverybench_scores} maps the row labels in Figure~\ref{fig:discoverybench_transfer}(b) to their task--host pairs and reports absolute final scores on the 0--100 scale. Pair 1 and Pair 2 share the same benchmark task but use different hosts; they are averaged before computing the task-level mean in panel (a).

DiscoveryBench evaluates generated hypotheses against target hypotheses using variable, relation, and context matching. We use its recorded \texttt{final\_score}, originally on a 0--1 scale, multiplied by 100. Means and paired differences are computed before rounding. Evaluation uses GPT-5.5 through Azure Chat Completions, separately from the research agents. This exploratory task subset was not randomly sampled and does not support a claim of representative benchmark coverage.

\begin{table}[htbp]
\centering
\caption{DiscoveryBench scores and row-label mapping for Figure~\ref{fig:discoverybench_transfer}(b).}
\label{tab:discoverybench_scores}
\small
\begin{tabularx}{\linewidth}{@{}l>{\raggedright\arraybackslash}Xlrrrr@{}}
\toprule
ID & Task & Host & N0 & R1 & R2 & R3 \\
\midrule
Pair 1 & nls\_ses / metadata\_0 & Agent Laboratory & 0.00 & 13.33 & 22.22 & 100.00 \\
Pair 2 & nls\_ses / metadata\_0 & EvoScientist & 12.50 & 44.44 & 44.44 & 0.00 \\
Pair 3 & nls\_incarceration / metadata\_2 & Agent Laboratory & 0.00 & 20.00 & 0.00 & 66.67 \\
Pair 4 & meta\_regression / metadata\_0 & Agent Laboratory & 9.38 & 12.00 & 25.00 & 0.00 \\
Pair 5 & archaeology / metadata\_13 & Agent Laboratory & 0.00 & 50.00 & 28.57 & 28.57 \\
Pair 6 & requirements\_engineering\_\newline for\_ML\_enabled\_systems / metadata\_0 & Agent Laboratory & 0.00 & 0.00 & 19.05 & 0.00 \\
\bottomrule
\end{tabularx}
\end{table}

\subsection{ResearchClawBench criterion-level scores}
\label{app:criterion_scores}

The table reports unweighted criterion scores and their weighted final scores,
consistent with the main analysis. Criterion descriptions are concise summaries of the supplied rubric
excerpts, not the complete benchmark rubric.

\begingroup
\small
\setlength{\tabcolsep}{1.5pt}
\renewcommand{\arraystretch}{1.04}
\setlength{\LTleft}{0pt}
\setlength{\LTright}{0pt}
\setlength{\LTcapwidth}{\linewidth}
\begin{longtable}{@{}p{0.485\linewidth}@{\hspace{0.03\linewidth}}p{0.485\linewidth}@{}}
\caption{Criterion-level scores and final scores, arranged in two columns.
AL denotes Agent Laboratory; Evo denotes EvoScientist.
Weights follow criterion order in each task header.
A dash denotes an unused criterion.}
\label{tab:criterion_scores}\\
\endfirsthead
\multicolumn{2}{@{}l}{\tablename~\thetable\ (continued)}\\[4pt]
\endhead
\multicolumn{2}{r@{}}{\footnotesize Continued on the next page}\\
\endfoot
\endlastfoot
\begin{minipage}[t]{\linewidth}\vspace{0pt}
\textbf{\mbox{Neuroscience-000}}\par
{\footnotesize Weights: 20\%, 20\%, 20\%, 20\%, 20\%.}\par\smallskip
\begin{tabular*}{\linewidth}{@{\extracolsep{\fill}}llrrrrrr@{}}
\toprule
Host & Cond. & C1 & C2 & C3 & C4 & C5 & Final \\
\midrule
AL & N0 & 12 & 5 & 0 & 0 & 28 & 9.00 \\
AL & R1 & 25 & 3 & 0 & 0 & 12 & 8.00 \\
AL & R2 & 5 & 0 & 0 & 0 & 12 & 3.40 \\
AL & R3 & 24 & 0 & 0 & 0 & 28 & 10.40 \\
Evo & N0 & 24 & 0 & 0 & 0 & 18 & 8.40 \\
Evo & R1 & 28 & 0 & 0 & 0 & 5 & 6.60 \\
Evo & R2 & 18 & 0 & 0 & 0 & 12 & 6.00 \\
Evo & R3 & 28 & 22 & 0 & 0 & 18 & 13.60 \\
ARK & N0 & 24 & 3 & 5 & 0 & 18 & 10.00 \\
ARK & R1 & 8 & 8 & 2 & 8 & 12 & 7.60 \\
ARK & R2 & 22 & 8 & 0 & 0 & 12 & 8.40 \\
ARK & R3 & 22 & 12 & 5 & 2 & 28 & 13.80 \\
\bottomrule
\end{tabular*}
\end{minipage}
&
\begin{minipage}[t]{\linewidth}\vspace{0pt}
\textbf{\mbox{Energy-001}}\par
{\footnotesize Weights: 10\%, 30\%, 20\%, 20\%, 20\%.}\par\smallskip
\begin{tabular*}{\linewidth}{@{\extracolsep{\fill}}llrrrrrr@{}}
\toprule
Host & Cond. & C1 & C2 & C3 & C4 & C5 & Final \\
\midrule
AL & N0 & 2 & 0 & 8 & 12 & 25 & 9.20 \\
AL & R1 & 15 & 8 & 3 & 2 & 28 & 10.50 \\
AL & R2 & 21 & 3 & 0 & 21 & 25 & 12.20 \\
AL & R3 & 3 & 0 & 0 & 0 & 28 & 5.90 \\
Evo & N0 & 24 & 25 & 8 & 3 & 28 & 17.70 \\
Evo & R1 & 35 & 12 & 12 & 18 & 34 & 19.90 \\
Evo & R2 & 8 & 28 & 37 & 18 & 28 & 25.80 \\
Evo & R3 & 85 & 22 & 18 & 18 & 31 & 28.50 \\
ARK & N0 & 45 & 38 & 18 & 18 & 31 & 29.30 \\
ARK & R1 & 35 & 35 & 32 & 8 & 28 & 27.60 \\
ARK & R2 & 25 & 42 & 34 & 22 & 28 & 31.90 \\
ARK & R3 & 25 & 42 & 18 & 22 & 24 & 27.90 \\
\bottomrule
\end{tabular*}
\end{minipage}
\\[10pt]
\begin{minipage}[t]{\linewidth}\vspace{0pt}
\textbf{\mbox{Life-001}}\par
{\footnotesize Weights: 35\%, 20\%, 15\%, 30\%.}\par\smallskip
\begin{tabular*}{\linewidth}{@{\extracolsep{\fill}}llrrrrrr@{}}
\toprule
Host & Cond. & C1 & C2 & C3 & C4 & C5 & Final \\
\midrule
AL & N0 & 2 & 0 & 0 & 2 & -- & 1.30 \\
AL & R1 & 3 & 0 & 12 & 6 & -- & 4.65 \\
AL & R2 & 12 & 3 & 12 & 1 & -- & 6.90 \\
AL & R3 & 2 & 2 & 12 & 0 & -- & 2.90 \\
Evo & N0 & 18 & 24 & 34 & 0 & -- & 16.20 \\
Evo & R1 & 18 & 12 & 32 & 0 & -- & 13.50 \\
Evo & R2 & 8 & 18 & 41 & 4 & -- & 13.75 \\
Evo & R3 & 18 & 18 & 28 & 0 & -- & 14.10 \\
ARK & N0 & 8 & 8 & 28 & 4 & -- & 9.80 \\
ARK & R1 & 8 & 12 & 28 & 0 & -- & 9.40 \\
ARK & R2 & 18 & 12 & 32 & 2 & -- & 14.10 \\
ARK & R3 & 12 & 8 & 31 & 0 & -- & 10.45 \\
\bottomrule
\end{tabular*}
\end{minipage}
&
\begin{minipage}[t]{\linewidth}\vspace{0pt}
\textbf{\mbox{Neuroscience-002}}\par
{\footnotesize Weights: 15\%, 25\%, 25\%, 20\%, 15\%.}\par\smallskip
\begin{tabular*}{\linewidth}{@{\extracolsep{\fill}}llrrrrrr@{}}
\toprule
Host & Cond. & C1 & C2 & C3 & C4 & C5 & Final \\
\midrule
AL & N0 & 2 & 12 & 0 & 0 & 0 & 3.30 \\
AL & R1 & 0 & 12 & 0 & 12 & 0 & 5.40 \\
AL & R2 & 0 & 12 & 0 & 12 & 2 & 5.70 \\
AL & R3 & 0 & 12 & 0 & 8 & 3 & 5.05 \\
Evo & N0 & 0 & 12 & 0 & 18 & 8 & 7.80 \\
Evo & R1 & 8 & 12 & 0 & 2 & 0 & 4.60 \\
Evo & R2 & 0 & 14 & 0 & 18 & 8 & 8.30 \\
Evo & R3 & 0 & 12 & 0 & 18 & 3 & 7.05 \\
ARK & N0 & 0 & 12 & 0 & 12 & 8 & 6.60 \\
ARK & R1 & 2 & 12 & 0 & 12 & 0 & 5.70 \\
ARK & R2 & 0 & 12 & 0 & 12 & 2 & 5.70 \\
ARK & R3 & 1 & 12 & 0 & 0 & 0 & 3.15 \\
\bottomrule
\end{tabular*}
\end{minipage}
\\[10pt]
\begin{minipage}[t]{\linewidth}\vspace{0pt}
\textbf{\mbox{Math-000}}\par
{\footnotesize Weights: 15\%, 15\%, 70\%.}\par\smallskip
\begin{tabular*}{\linewidth}{@{\extracolsep{\fill}}llrrrrrr@{}}
\toprule
Host & Cond. & C1 & C2 & C3 & C4 & C5 & Final \\
\midrule
AL & N0 & 45 & 32 & 8 & -- & -- & 17.15 \\
AL & R1 & 32 & 38 & 12 & -- & -- & 18.90 \\
AL & R2 & 44 & 36 & 15 & -- & -- & 22.50 \\
AL & R3 & 34 & 32 & 12 & -- & -- & 18.30 \\
Evo & N0 & 24 & 27 & 12 & -- & -- & 16.05 \\
Evo & R1 & 3 & 5 & 12 & -- & -- & 9.60 \\
Evo & R2 & 24 & 36 & 18 & -- & -- & 21.60 \\
Evo & R3 & 28 & 37 & 18 & -- & -- & 22.35 \\
\bottomrule
\end{tabular*}
\end{minipage}
&
\begin{minipage}[t]{\linewidth}\vspace{0pt}
\textbf{\mbox{Math-001}}\par
{\footnotesize Weights: 40\%, 30\%, 30\%.}\par\smallskip
\begin{tabular*}{\linewidth}{@{\extracolsep{\fill}}llrrrrrr@{}}
\toprule
Host & Cond. & C1 & C2 & C3 & C4 & C5 & Final \\
\midrule
AL & N0 & 0 & 8 & 29 & -- & -- & 11.10 \\
AL & R1 & 0 & 46 & 31 & -- & -- & 23.10 \\
AL & R2 & 8 & 39 & 42 & -- & -- & 27.50 \\
AL & R3 & 23 & 38 & 31 & -- & -- & 29.90 \\
Evo & N0 & 6 & 22 & 26 & -- & -- & 16.80 \\
Evo & R1 & 35 & 34 & 28 & -- & -- & 32.60 \\
Evo & R2 & 32 & 45 & 33 & -- & -- & 36.20 \\
Evo & R3 & 35 & 43 & 36 & -- & -- & 37.70 \\
\bottomrule
\end{tabular*}
\end{minipage}
\\[10pt]
\begin{minipage}[t]{\linewidth}\vspace{0pt}
\textbf{\mbox{Math-002}}\par
{\footnotesize Weights: 40\%, 30\%, 30\%.}\par\smallskip
\begin{tabular*}{\linewidth}{@{\extracolsep{\fill}}llrrrrrr@{}}
\toprule
Host & Cond. & C1 & C2 & C3 & C4 & C5 & Final \\
\midrule
AL & N0 & 0 & 0 & 0 & -- & -- & 0.00 \\
AL & R1 & 3 & 5 & 3 & -- & -- & 3.60 \\
AL & R2 & 12 & 0 & 12 & -- & -- & 8.40 \\
AL & R3 & 12 & 0 & 0 & -- & -- & 4.80 \\
Evo & N0 & 7 & 8 & 18 & -- & -- & 10.60 \\
Evo & R1 & 5 & 6 & 5 & -- & -- & 5.30 \\
Evo & R2 & 14 & 3 & 15 & -- & -- & 11.00 \\
Evo & R3 & 5 & 2 & 2 & -- & -- & 3.20 \\
\bottomrule
\end{tabular*}
\end{minipage}
&
\begin{minipage}[t]{\linewidth}\vspace{0pt}
\textbf{\mbox{Math-003}}\par
{\footnotesize Weights: 40\%, 35\%, 25\%.}\par\smallskip
\begin{tabular*}{\linewidth}{@{\extracolsep{\fill}}llrrrrrr@{}}
\toprule
Host & Cond. & C1 & C2 & C3 & C4 & C5 & Final \\
\midrule
AL & N0 & 0 & 0 & 0 & -- & -- & 0.00 \\
AL & R1 & 12 & 15 & 0 & -- & -- & 10.05 \\
AL & R2 & 6 & 12 & 0 & -- & -- & 6.60 \\
AL & R3 & 12 & 12 & 0 & -- & -- & 9.00 \\
Evo & N0 & 31 & 9 & 3 & -- & -- & 16.30 \\
Evo & R1 & 21 & 3 & 0 & -- & -- & 9.45 \\
Evo & R2 & 15 & 15 & 0 & -- & -- & 11.25 \\
Evo & R3 & 18 & 5 & 3 & -- & -- & 9.70 \\
\bottomrule
\end{tabular*}
\end{minipage}
\\[10pt]
\begin{minipage}[t]{\linewidth}\vspace{0pt}
\textbf{\mbox{Physics-002}}\par
{\footnotesize Weights: 20\%, 20\%, 15\%, 20\%, 25\%.}\par\smallskip
\begin{tabular*}{\linewidth}{@{\extracolsep{\fill}}llrrrrrr@{}}
\toprule
Host & Cond. & C1 & C2 & C3 & C4 & C5 & Final \\
\midrule
AL & N0 & 0 & 0 & 0 & 0 & 0 & 0.00 \\
AL & R1 & 5 & 12 & 2 & 12 & 2 & 6.60 \\
AL & R2 & 18 & 32 & 3 & 28 & 34 & 24.55 \\
AL & R3 & 2 & 3 & 2 & 15 & 0 & 4.30 \\
Evo & N0 & 32 & 45 & 28 & 38 & 48 & 39.20 \\
Evo & R1 & 6 & 8 & 5 & 28 & 5 & 10.40 \\
Evo & R2 & 32 & 36 & 24 & 32 & 78 & 43.10 \\
Evo & R3 & 12 & 25 & 8 & 8 & 25 & 16.45 \\
ARK & N0 & 28 & 35 & 24 & 12 & 36 & 27.60 \\
ARK & R1 & 38 & 42 & 34 & 18 & 48 & 36.70 \\
ARK & R2 & 31 & 32 & 22 & 28 & 42 & 32.00 \\
ARK & R3 & 32 & 42 & 28 & 18 & 48 & 34.60 \\
\bottomrule
\end{tabular*}
\end{minipage}
&
\begin{minipage}[t]{\linewidth}\vspace{0pt}
\textbf{\mbox{Information-003}}\par
{\footnotesize Weights: 15\%, 20\%, 30\%, 20\%, 15\%.}\par\smallskip
\begin{tabular*}{\linewidth}{@{\extracolsep{\fill}}llrrrrrr@{}}
\toprule
Host & Cond. & C1 & C2 & C3 & C4 & C5 & Final \\
\midrule
AL & N0 & 8 & 8 & 0 & 4 & 0 & 3.60 \\
AL & R1 & 15 & 8 & 0 & 8 & 0 & 5.45 \\
AL & R2 & 8 & 0 & 6 & 0 & 0 & 3.00 \\
AL & R3 & 12 & 6 & 12 & 8 & 0 & 8.20 \\
Evo & N0 & 8 & 22 & 12 & 8 & 0 & 10.80 \\
Evo & R1 & 18 & 6 & 12 & 6 & 0 & 8.70 \\
Evo & R2 & 12 & 22 & 3 & 6 & 0 & 8.30 \\
Evo & R3 & 14 & 3 & 12 & 3 & 0 & 6.90 \\
\bottomrule
\end{tabular*}
\end{minipage}
\\[10pt]
\end{longtable}
\endgroup

\subsection{ResearchClawBench criterion summaries}
\label{app:criterion_summaries}

The descriptions below summarize what each criterion assesses.
They describe benchmark targets, not findings established by the
evaluated agents.

\paragraph{\mbox{Neuroscience-000}.}
C1: attack-classifier precision--recall performance across conditions.
C2: cross-laboratory SHAP comparisons.
C3: male--female SHAP comparisons.
C4: RI--CSDS environment SHAP comparisons.
C5: permutation-based feature importance.

\paragraph{\mbox{Energy-001}.}
C1: construction of the Scottish wind-curtailment power-system model.
C2: constrained versus unconstrained optimal power flow.
C3: dispatched and curtailed wind over time.
C4: loading of the Scotland--England boundary link.
C5: the open-source PyPSA-GB representation of the British power system.

\paragraph{\mbox{Life-001}.}
C1: immune-response distributions for optimized vaccines.
C2: coverage across response thresholds.
C3: optimization runtime scaling.
C4: recall of experimentally validated neoantigens against
ranking-based comparators.

\paragraph{\mbox{Neuroscience-002}.}
C1: FlyTracing dataset construction.
C2: fusion of morphological and image features.
C3: EmbedNet weighting ablations.
C4: robustness to imaging artifacts.
C5: visualization of neuron-segment embeddings.

\paragraph{\mbox{Math-000}.}
C1: pseudo-depth estimation from image geometry.
C2: depth cascade matching using depth-partitioned association.
C3: SparseTrack tracking performance on MOT17, MOT20, and DanceTrack.

\paragraph{\mbox{Math-001}.}
C1: accelerated convergence on a high-dimensional Lasso problem.
C2: convergence trajectories and stability-related evidence.
C3: robustness on an ill-conditioned regression problem.

\paragraph{\mbox{Math-002}.}
C1: multi-agent path-finding performance in high-density settings.
C2: remaining collisions and solution cost.
C3: zero-shot generalization across map structures.

\paragraph{\mbox{Math-003}.}
C1: geometry-proof performance on IMO-AG-30.
C2: synthetic training-data scale.
C3: traceback-based identification of an unused premise and
the resulting theorem generalization.

\paragraph{\mbox{Physics-002}.}
C1: fidelity scaling with qubit count at fixed depth.
C2: fidelity scaling with depth for 40 qubits.
C3: depth-dependent fidelity estimates for 56 qubits.
C4: gate-counting predictions and uncertainty.
C5: counts-weighted linear XEB reproduction on the
40-qubit verification subset.

\paragraph{\mbox{Information-003}.}
C1: evidence connecting feature entanglement with detection failures.
C2: the DIDS-MFL disentanglement and graph-diffusion framework.
C3: few-shot intrusion-detection performance across benchmarks.
C4: component ablations and visualizations.
C5: comparison with large language models.

\section{Detailed Case Studies of RAC Coordination}
\label{app:rac_trace_cases}

We examine five trajectories to illustrate how RAC changes the work performed between agent boundaries. The cases cover direct reassignment of corrective work, artifact-grounded verification without a change of agent order, improved task relevance under communication-enabled execution, and a failure to translate local completion into global progress. We emphasize the concrete intervention visible in each trace: which defect was identified, which capability acted next, what changed on disk, and what the external judge subsequently assessed.

\paragraph{Evidence and interpretation.}
These are selected qualitative cases, not additional controlled ablations. The ARK records are exported hop summaries paired with artifacts and external scoring receipts. The Math cases are identified by individual episode records. Numerical outcomes refer to the final scores reported in Table~\ref{tab:task_scores}; older receipts are not used to decompose those scores. Their differences describe the N0 and RAC executions, rather than isolated effects of a coordination component. All hop indices are zero-based. Quoted passages below reproduce short excerpts, with Markdown styling removed where needed; the accompanying descriptions are our reconstruction from the records, not additional agent dialogue. We distinguish the \emph{host reviewer}, which critiques research outputs, the \emph{RAC verifier}, which checks work-contract evidence, and the \emph{external judge}, which applies the benchmark criteria. We have not independently rerun the scientific computations.

\subsection{ARK on \mbox{Life-001}: direct corrective handoffs}
\label{app:case_ark_life}

\paragraph{Outcome and coordination change.}
The useful change in this R2 trajectory was a shorter path from a concrete review finding to a revised artifact. Whereas the recorded native successor after review was \texttt{planner}, RAC repeatedly selected \texttt{writer} directly. This capability did more than edit prose: it reported modifying analysis code, regenerating comparisons, and checking claims against the underlying data. Thus, the observed benefit was not the introduction of review---ARK already had a native review loop---but a change in who acted immediately on its findings.

\begin{tracebox}{Trace: repairing the comparison being reported.}
\begin{enumerate}
\item \textbf{Hop 3, reviewer: identify a method--result mismatch.} The reviewer reported that the headline coverage curve used the greedy baseline rather than the proposed MinSum objective. It also identified missing threshold sensitivity and unsupported generalization from a single simulated tumor. The recorded successor was \texttt{planner}.
\begin{sourcebox}
``E1 coverage curve uses greedy, not MinSum''
\end{sourcebox}
\item \textbf{Hop 4, writer: act directly on the implicated artifacts.} The actual next capability was \texttt{writer}. It reported adding \texttt{compute\_greedy\_minsum\_curve()} to the experiment code and regenerating a figure with both curves, rather than merely renaming the existing result. The export recorded 15 scientific-artifact changes at this hop.
\begin{sourcebox}
``regenerated figure with both curves; rewrote E1 text to position greedy as baseline and MinSum as proposed method''
\end{sourcebox}
\item \textbf{Hop 5, reviewer: check the repair.} The reviewer explicitly recognized the added MinSum curve, threshold analysis, corrected coverage numbers, and narrower claims. It then requested that replicate variability be made visible and that clinical-feasibility language be qualified.
\item \textbf{Hop 6, writer: complete the next local correction.} RAC again selected the writer directly. The report acquired the replicate-level range of mean response probabilities, and claims of clinical feasibility were replaced with claims about computational tractability at the tested scale.
\end{enumerate}

The final report corroborates the central repair: it contains separate greedy and MinSum coverage curves, with MinSum coverage of 97.9\% at budget four and 99.2\% at budget ten. It also reports coverage at multiple response thresholds rather than relying entirely on a single cutoff. These persisted changes connect the selected capability to the artifact that the reviewer had challenged.
\end{tracebox}

\begin{tracebox}{Trace: replacing an unsupported explanation with a data check.}
A later handoff shows that the same pattern could improve interpretation as well as presentation. At hop 11, the reviewer challenged the explanation for low-response cells. At hop 12, the writer inspected per-peptide scores and reported that five of the eight cells had near-zero scores for every candidate, while the other three had their highest scores among already selected elements. Its summary stated:
\begin{sourcebox}
``the claim was not just unsupported'' \ldots\ ``contradicted by the data''
\end{sourcebox}
The revision replaced speculation about excluded vaccine elements with a description grounded in those observations. This is a local evidence--action closure: feedback specified a claim, the next capability checked the relevant data, and the explanation changed.
\end{tracebox}

\paragraph{Relation to the reported outcome.}
Table~\ref{tab:task_scores} reports final scores of 9.80 for N0 and 14.10 for R2, a difference of 4.30 points. The trace supports the local repairs described above, not a quantitative attribution of the final score gain to individual handoffs.

The case nevertheless provides a concrete positive example of RAC changing the native successor and closing local defects. Its boundary is equally specific: efficacy experiments still used a single simulated tumor with 100 cells per replicate, rather than the required seven-patient, 10,000-cell analysis. The run also continued through repeated revisions to the 46-hop limit. Direct corrective handoffs improved a partially aligned study without completing the benchmark's broader scientific requirements.

\paragraph{Delivered-paper evidence: the correction survives into the manuscript.}
The final paper explicitly separates the proposed objective from the baseline and qualifies its comparative result:
\begin{sourcebox}
``Under the proposed MinSum objective, coverage rises from 53.2\% [50.1\%, 56.3\%] at $k{=}1$ to 94.0\% [92.3\%, 95.4\%] at $k{=}3$, reaches 97.9\% [96.8\%, 98.7\%] at $k{=}4$, and plateaus at 99.2\% [98.4\%, 99.7\%] by $k{=}7$.''

``The MinSum point-estimate curve is higher than greedy at every budget $k \geq 3$ (though CI overlap at $k{=}3$ and the non-significant E5 comparison at $p{=}0.0625$ caution against claims of strict superiority).''
\end{sourcebox}
This is more informative than a generic claim that a figure was fixed: the delivered text identifies which curve represents MinSum and distinguishes point-estimate differences from statistical superiority. The paper also bounds the experiment:
\begin{sourcebox}
``Vaccine efficacy experiments (E1--E3, E5) use a single simulated tumor model with $M=12$ mutations and $N=100$ cells per replicate. Runtime scaling (E4) spans 7 independently parameterized instances to provide cross-instance evidence for computational cost.''
\end{sourcebox}
The distinction prevents seven runtime instances from being read as seven independent efficacy studies. These passages support the positive interpretation of the trace: corrective handoffs improved the fidelity and qualification of the delivered comparison, even though they did not supply the missing patient-scale evaluation.

\paragraph{Residual inconsistency: repaired delivery does not certify the objective.}
The same paper defines a minimization objective that decomposes as
\[
\mathcal{L}_{\mathrm{min}}(S)
=\frac{1}{N}\sum_{i=1}^{N}\log(1-P_i(S))
=\sum_{j\in S}w_j,
\qquad
w_j=\frac{1}{N}\sum_{i=1}^{N}\log(1-p_{ij}),
\]
but its accompanying instruction reads:
\begin{sourcebox}
``The optimal $k$-element solution therefore reduces to ranking elements by $w_j$ (descending) and selecting the top-$k$.''
\end{sourcebox}
For the stated minimization of a sum of fixed weights, the selected weights should instead be the smallest ones. This establishes a contradiction in the paper's description, not that the implementation necessarily used the wrong order. The paper also acknowledges that this is ``structurally the same ranking paradigm the Introduction critiques''. Consequently, the corrected comparison is evidence of improved reporting, not evidence of a newly established combinatorial optimization paradigm. RAC's observed success here is local: it closes particular reviewer-identified defects while leaving a mathematical description that warrants further checking.

\subsection{ARK on \mbox{Physics-002}: repair followed by cross-capability checking}
\label{app:case_ark_physics}

\paragraph{Outcome and coordination change.}
The selected R3 snapshot scored 34.6, compared with 27.6 for N0. RAC departed from the recorded native successor at several boundaries and assigned concrete defects across writer, coder, and experimenter capabilities. The resulting trajectory is useful because it includes both a correction to existing evidence and a second check that caught an error introduced while extending a figure.

\begin{tracebox}{Trace: reconciling reported values with computed results.}
\begin{enumerate}
\item \textbf{Hop 7, reviewer: localize the discrepancy.} The reviewer identified XEB table values that disagreed with the experiment manifest, a missing N56 result table, undisclosed differences in shot counts, and incorrect significance statements.
\begin{sourcebox}
``d=10: paper reports 0.557 vs manifest 0.4502''
\end{sourcebox}
The recorded successor was \texttt{planner}.
\item \textbf{Hop 8, writer: revise before replanning.} RAC selected \texttt{writer} instead. The writer added the N56 table, disclosed 100 shots per instance at depth 24 versus 20 at other depths, and corrected a caption that described a nonexistent second axis.
\item \textbf{Hop 9, coder: inspect the partially repaired report.} Rather than proceeding directly to the writer's proposed reviewer successor, execution moved to \texttt{coder}. It verified the seven XEB depth values and found a residual discrepancy between the discussion and conclusion:
\begin{sourcebox}
``The Discussion section already correctly stated `23$\sigma$' --- the Conclusion was the only place with the stale `26$\sigma$'.''
\end{sourcebox}
\item \textbf{Hop 10, experimenter: check against numerical artifacts.} The experimenter checked the XEB values against the instance records and reported a maximum discrepancy of 0.00028, within the stated rounding tolerance. It also corrected Transport confidence intervals and added the corresponding N56 table.
\end{enumerate}

The final report contains the corrected depth-10 XEB value of 0.450. The significance statements in these excerpts remain agent calculations, not independent validation by our audit; the directly checked fact is that the handoff led to reconciliation of the persisted report and its cited outputs.
\end{tracebox}

\begin{tracebox}{Trace: detecting a bug in a newly added figure.}
At hop 42, the coder handled a figure-specific issue and expanded the summary figure from two panels to three, adding the N56 MB depth scan. At hop 43, the experimenter checked that addition and identified a denominator error in the pooled confidence intervals:
\begin{sourcebox}
``The panel (c) code hardcoded \texttt{n\_total = n * 20}''
\end{sourcebox}
The depth-24 data used 100 shots per instance, not 20. The experimenter reported changing the interval from $[0.1014,0.1428]$ to $[0.1119,0.1302]$. The final report both references panel (c) and records the rounded corrected interval $[0.112,0.130]$, together with 100 shots per instance and 5,000 total trials. This sequence links a scoped figure task to implementation, a distinct checking step, and a corrected delivered result.
\end{tracebox}

\paragraph{RAC's contribution and its scope.}
The archived provenance for this ARK R3 case identifies an expanded-budget continuation. We therefore use its trace as qualitative evidence of local repairs, not as a matched-budget estimate of their benefit.
Unlike a fixed traversal of role names, the observed routing followed the type of outstanding work: document reconciliation, code and figure repair, and numerical checking. Typed issue identifiers in the summaries made the requested changes concrete. The cross-capability check was productive because it found a specific new bug, not merely because a second agent declared the work complete. The export does not include the underlying RAC verifier ledger for these ARK hops, so this example establishes a corrective handoff pattern rather than independently verified contract compliance at every boundary.

N0's report had introduced an additional ``unbiased XEB'' rescaling, whereas R3 retained values on the target scale. This between-run method difference must not be mistaken for a logged repair of N0's code by R3. The within-R3 evidence above instead shows how concrete inconsistencies were repaired and checked before delivery; it does not decompose the 7.0-point final-score difference. Gate-counting uncertainty bands and the required mirror-circuit inference remained incomplete.

\paragraph{Delivered-paper evidence: a correct patch and an obsolete claim coexist.}
The final manuscript provides unusually direct evidence of both the value and the limits of local repair. Its comparison paragraph retains the following sentence:
\begin{sourcebox}
``The largest within-CI spread at any single depth is at d=12 ($F_{\mathrm{XEB}}=0.457$ vs. $F_{\mathrm{MB}}=0.476$, CIs [0.376,0.538] and [0.445,0.507] respectively).''
\end{sourcebox}
Yet the same paragraph subsequently states:
\begin{sourcebox}
``The largest between-method differences occur at d=10 ($F_{\mathrm{XEB}}=0.450$ vs. $F_{\mathrm{MB}}=0.518$, $\Delta=0.068$) and d=16 ($F_{\mathrm{XEB}}=0.368$ vs. $F_{\mathrm{MB}}=0.305$, $\Delta=0.063$).''
\end{sourcebox}
The table supports the latter values: the depth-12 difference is only 0.019, compared with 0.068 and 0.063 at depths 10 and 16. The corresponding intervals also overlap at those depths, so the ``within-CI'' qualification does not resolve the obsolete superlative. Both sentences occur in the original \texttt{main.tex}, not only in the Markdown export. A corrected statement was delivered without fully reconciling the surrounding text.

The figure repair is more cleanly reflected in the final artifact. The paper reports the depth-24 pooled interval as $[0.112,0.130]$, and its table lists 100 shots per instance and 5,000 total trials. These are the quantities implicated in the hop-43 denominator correction. Thus, the paper corroborates a successful numerical patch while exposing a separate synchronization failure. The mechanism-level lesson is not that cross-capability checking failed altogether, but that checking a named value or figure can succeed without checking every dependent statement in the manuscript. The higher score and the remaining contradiction can therefore coexist.

\subsection{EvoScientist on \mbox{Math-001}: explicit delivery verification with unchanged agent order}
\label{app:case_evo_math001}

\paragraph{Outcome and coordination change.}
The selected R3 episode scored 37.7, compared with 16.8 for N0. Its executed sequence remained planner, research, code, debug, data analysis, and writing. This case therefore illustrates a different RAC contribution: a work contract distinguished an agent's nonempty response from an actual implementation, even though the next agent retained the native position in the workflow.

\begin{tracebox}{Trace: a nonempty response fails the delivery contract.}
\begin{enumerate}
\item \textbf{Hop 2, code contract.} The decision required changed code/result artifacts and nonempty output. These requirements were recorded separately, rather than treating a successful return as evidence of implementation.
\item \textbf{Hop 2, code invocation.} The agent announced that it was about to delegate, but did not deliver the required implementation:
\begin{sourcebox}
``Good. Both /code/ and /outputs/ are empty --- we're starting from scratch.''
\end{sourcebox}
\item \textbf{Hop 2, RAC verifier.} The actual evaluation event recorded \texttt{nonempty\_output: passed}, \texttt{artifact\_changed: failed}, and verdict \texttt{refuted}:
\begin{sourcebox}
``verification advisory (refuted): one or more contract evidence checks failed''
\end{sourcebox}
\item \textbf{Hop 3, debug invocation.} The next capability delivered the missing implementation rather than only commenting on it. Its output began:
\begin{sourcebox}
``Implementation complete.'' \ldots\ ``Code \& Results Delivered''
\end{sourcebox}
The artifact record includes solver modules, a Lasso experiment, convergence plots, and method-comparison results. The subsequent verifier event recorded \texttt{supported} and listed the changed code and output files.
\end{enumerate}
\end{tracebox}

\paragraph{Why this is a useful positive contract example.}
RAC added an explicit, machine-recorded distinction between \emph{intention to perform work} and \emph{evidence that the work was delivered}. That distinction exposed a gap that a nonempty-output check alone would have missed. The following native debug capability then filled the gap, after which analysis and writing produced a substantive report. Thus, a visibly unchanged agent order did not imply unchanged coordination semantics: the boundary now carried an auditable delivery verdict.

The scope of the inference matters. The code invocation had already proposed debug as its successor; the trace does not show a verifier-induced reroute, forced retry, or rollback. Nor does this single trajectory isolate whether debug would have repaired the missing implementation without the advisory. The supported observation is the sequence of explicit detection followed by artifact-backed recovery, which is consistent with a beneficial role for contracts and verification without changing the agent roster or order.

\paragraph{Relation to the result.}
The N0 receipt describes a VOS method that underperformed its baseline; the R3 submission provided a more complete implementation, comparator analysis, convergence evidence, and a 460-line report. These differences explain why the external judge had more relevant evidence to assess, while the trace identifies a concrete delivery gap successfully closed along the way. The result was still partial: the run exceeded its input-token budget after writing, before a native review invocation, and did not fully establish the requested convergence target. The 20.9-point difference is therefore an outcome associated with this recovered trajectory, not a measured causal effect of the verifier alone.

\paragraph{Delivered-paper evidence: a working comparator, not a new winning algorithm.}
The generated paper explicitly characterizes the implemented method as matching an established baseline:
\begin{sourcebox}
``VOS matches FISTA: At $r=3$, the VOS discretization is algorithmically equivalent to FISTA with gradient restart. The identical convergence trajectories (92 iterations, rates within 0.01 of each other) confirm that the VOS framework correctly reproduces Nesterov acceleration.''
\end{sourcebox}
Its comparison table gives both FISTA and VOS 92 iterations, with fitted rates of 6.60 and 6.61, respectively. This grounds the positive case more precisely: the recovered implementation supplies a functioning accelerated baseline and relevant comparison evidence. It does not demonstrate a distinct algorithm outperforming FISTA. That narrower interpretation is consistent with the delivery contract: the missing implementation was supplied, but the verifier did not establish algorithmic novelty.

\paragraph{Residual numerical and theoretical overreach.}
Two passages show why artifact-backed delivery must be distinguished from validation of the scientific conclusions. First, the paper states:
\begin{sourcebox}
``All three methods converge to essentially the same point (objective within $3 \times 10^{-9}$ of each other).''
\end{sourcebox}
The comparison table instead lists final suboptimalities of $1.40\times10^{-5}$ for FISTA, $1.40\times10^{-6}$ for ADMM, and $1.37\times10^{-5}$ for VOS. If these use a common reference optimum, the FISTA--ADMM objective difference is approximately $1.26\times10^{-5}$, inconsistent with the quoted tolerance. If different references were used, the paper would need to explain how the cross-method claim was obtained. We flag an unreconciled reporting discrepancy rather than claim to have recomputed the objectives.

Second, the paper's Lyapunov diagnostic uses changing reference values:
\begin{sourcebox}
``where $f^\star$ and $x^\star$ are dynamically updated to the best observed values.''

``Non-monotonic steps: 13 out of 92 (14.1\%)''
\end{sourcebox}
The discussion acknowledges that ``the dynamic update of $f^\star$ and $x^\star$ introduces reference-point shifts'', but the conclusion nevertheless describes the experiment as ``exhibiting the predicted Lyapunov monotonicity (85.9\% monotonic steps, $p = 2.87$ decay exponent)''. A moving-reference, partly non-monotone diagnostic is not a direct verification of a fixed-reference monotonicity theorem. Nor do these discrete observations by themselves refute a continuous-time theorem. The appropriate claim is evidence of empirical decay under the implemented diagnostic. This separates RAC's observable contribution---detecting and closing a delivery gap---from stronger mathematical claims that its artifact checks did not validate.

\subsection{Agent Laboratory on \mbox{Math-003}: greater task relevance under communication-enabled execution}
\label{app:case_agentlab_math003}

\paragraph{Outcome and coordination condition.}
The selected R1 episode scored 10.05, compared with 0 for N0. N0 studied an unrelated synthetic-transformation setup. R1 instead developed a geometry-proof experiment with self-play and comparison baselines. Because R1 preserved the native successor and did not enable work-contract verification, this case examines improvement under communication-enabled execution rather than dynamic selection or verifier-guided repair.

\begin{tracebox}{Trace: carrying a task-relevant plan into experiments.}
\begin{enumerate}
\item \textbf{Runtime decisions.} The recorded decision reason was:
\begin{sourcebox}
``native fixed successor with SharedNet communication''
\end{sourcebox}
The workflow retained literature review, plan formulation, data preparation, experiments, interpretation, and report writing. Evaluation events explicitly stated that verification was disabled.
\item \textbf{Hop 1, plan formulation.} The planning dialogue framed proof steps as actions in a geometry-solving system, combining sequence generation, search, and self-play. Its stated goal was:
\begin{sourcebox}
``autonomously proving olympiad-level geometry theorems without human demonstrations''
\end{sourcebox}
\item \textbf{Hop 3, experiments.} Execution attempted the task-relevant experimental plan, encountering several 600-second code-execution timeouts and native repair attempts before producing results. These repairs belong to the host's internal execution machinery, not to an R3 verifier.
\item \textbf{Hop 4, interpretation.} The next stage discussed the actual failures of that experiment rather than substituting a different research question:
\begin{sourcebox}
``0/10 on the random walk baseline, 0/10 on the supervised imitation learning baseline, and 0/10 on our SPNS-EPS self-play approach''
\end{sourcebox}
\item \textbf{Hop 5, writing.} A report of the geometry experiment was delivered. Input-token exhaustion prevented the subsequent refinement stage.
\end{enumerate}
\end{tracebox}

\paragraph{The communication-related interpretation.}
The positive change was continuity of a relevant research objective across planning, implementation, and interpretation. Under the communication-enabled condition, the host produced a connected, albeit unsuccessful, geometry experiment instead of an off-task study. This motivates communication as a way of improving what successive agents coordinate around without changing their order. However, the recorded host dialogue is not itself proof of SharedNet message consumption. The available evidence does not link a particular SharedNet message to a downstream decision, and R1 differs from N0 in its integration context as well as communication availability. We therefore treat this as a communication-consistent example, not a component-level causal demonstration.

The score increase should also be read as improved relevance, not improved theorem-proving performance: every reported method solved 0 of the ten evaluated problems, and the study did not reproduce the full target benchmark. Planning consumed approximately 70.5\% of the episode's recorded input tokens. The trajectory shows the value of preserving a meaningful task across handoffs, while also showing that relevance alone does not ensure successful execution or efficient use of the budget.

\paragraph{Delivered-paper evidence: relevant and candid negative results.}
The R1 manuscript reports the failure directly rather than manufacturing successful proof results:
\begin{sourcebox}
``Our experiments reveal a stark negative result: all three approaches achieve exactly 0/10 (0.0\%) proof success on the IMO test set. The SPNS-EPS self-play agent achieves an immediate $+1.000$ reward at episode 0 and maintains this reward with identically zero policy loss and zero value loss across all 120 training episodes.''
\end{sourcebox}
It positions its contribution accordingly:
\begin{sourcebox}
``The primary contribution of this work is not a novel method that achieves state-of-the-art performance, but rather an empirical characterization of the failure modes''
\end{sourcebox}
The latter is an excerpt from a longer sentence. Together, these passages reinforce the positive but limited interpretation: R1 delivered an on-topic experiment and openly reported unsuccessful outcomes. The relevance gain does not depend on accepting the manuscript's explanation of those outcomes.

\paragraph{Code evidence: the failure explanation misses implementation-level causes.}
The paper interprets a zero supervised loss as successful memorization:
\begin{sourcebox}
``A loss of exactly zero indicates that the network assigns probability 1.0 to the target action index for every training example, achieving perfect memorization of the training set.''
\end{sourcebox}
However, the delivered training loop constructs its legal-action mask from the target action alone. The following contiguous excerpt is from \texttt{run\_experiments.py}:
\begin{sourcebox}
\begin{verbatim}
mask, _ = build_legal_mask([target_action])
target_idx = action_to_idx(target_action)
mask[0, target_idx] = 1.0
_, logits = sup_net([gdata], goal_t, mask)
loss = F.cross_entropy(logits, torch.tensor([target_idx]))
\end{verbatim}
\end{sourcebox}
The forward pass masks other logits, and the loss is cross-entropy against that same target. With only the target class left available, probability one and a near-zero loss follow from the mask; they are not evidence that the network learned to distinguish competing actions. This static implementation evidence supports a more specific explanation than memorization.

The immediate self-play reward has a similarly concrete cause. In \texttt{random\_provable\_problem()}, the generator repeatedly applies a theorem and updates \texttt{state = ns}. It then sets the goal to the conclusion of the last applied theorem and returns that already-expanded state, rather than the original premises. The decisive terminal statements are:
\begin{sourcebox}
\begin{verbatim}
state.goal = goal
return state
\end{verbatim}
\end{sourcebox}
The theorem application has already inserted its conclusion into the state's relations; the goal check tests membership in those relations. When there is no recorded application, the alternative branch chooses an existing relation as the goal. Thus, the returned training instance already contains its goal. This explains why immediate success can occur without a learned proof policy. The paper instead discusses forward chaining as producing goals reachable ``in zero or one steps'', framing the problem primarily as an overly easy curriculum.

These are static code findings, not results of a rerun or proof that fixing the two issues would solve the benchmark. They identify a missed distinction between a genuinely informative negative experiment and an implementation artifact. R1 improved task continuity and reporting relevance, but did not provide the semantic checking needed to catch these defects before they became an explanatory narrative. Since verification was disabled in R1, this is not a failed R3 verifier verdict.

\subsection{EvoScientist on \mbox{Math-003}: local contracts can leave the global objective unfinished}
\label{app:case_evo_math003}

\paragraph{Outcome and failure boundary.}
The selected R3 episode scored 9.7, below the N0 score of 16.3. This is not a case in which RAC broke an otherwise successful prover: N0 also reported 0/30 solved problems. The useful contrast is that R3 developed and validated a narrower intermediate artifact, then proceeded to analyze and write about it while the planned system remained incomplete.

\begin{tracebox}{Trace: stage completion substitutes for system completion.}
\begin{enumerate}
\item \textbf{Hop 0, planner.} The plan specified six implementation stages, including a neural policy/value component and Monte Carlo tree search for geometry proving.
\item \textbf{Hop 2, code.} The implementation stopped at the parser and deduction-engine foundation:
\begin{sourcebox}
``Stage 0 Complete: Parser \& Deduction Engine''
\end{sourcebox}
It reported 89 passing tests and successful parsing of all 30 problems. The contract's executable evidence checks required changed code/result artifacts and nonempty output; the RAC verifier returned \texttt{supported}.
\item \textbf{Hop 3, debug.} Debugging fixed five reported bugs and increased the test count to 102. These were genuine local improvements to the implemented foundation.
\item \textbf{Hop 4, data analysis.} Instead of returning to the unimplemented neural and search components, execution moved to analysis of the existing artifact: relation counts, predicate distributions, and related diagnostics, with 11 generated figures. Its contract also received \texttt{supported}.
\item \textbf{Hops 5--6, writing and host review.} The writer produced an 861-line report centered on Stage 0. The host reviewer gave 8/10 after checking such properties as the passing tests and the reported 0/30 result. All seven RAC verification events in the episode were \texttt{supported}.
\end{enumerate}
\end{tracebox}

\paragraph{What the negative case says about RAC.}
The verification results were compatible with their local evidence rules: files changed, outputs were nonempty, and the parser was tested. The weakness was the connection between those local checks and the remaining scientific obligations. The contract objective named autonomous geometry solving, but its executable criteria did not establish that the full planned solver had been implemented. Runtime selection continued along the native sequence, allowing analysis and writing to consolidate the partial system rather than return to the missing stages.

This illustrates a failure mode in which locally supported work accumulates without closing the global objective. RAC made the intermediate work explicit and checkable, but did not prevent premature progression from a useful foundation to a final report. The episode ultimately exceeded its input-token limit; nevertheless, the narrowing to Stage 0 was already visible before termination. Budget exhaustion alone therefore does not explain the observed sequence. The case motivates connecting contracts and subsequent selection to unresolved end-to-end requirements, rather than interpreting \texttt{supported} as scientific completion. It does not establish that contracts necessarily caused the narrowing or generally reduce performance.

\paragraph{Delivered-paper evidence: the missing system becomes future work.}
The final manuscript makes the gap between local completion and the research objective explicit. It reports:
\begin{sourcebox}
``The most important empirical result: 0/30 problems are solvable by forward chaining alone.''
\end{sourcebox}
Elsewhere, it describes its foundation in stronger terms:
\begin{sourcebox}
``Our implementation---parser, definition expander, rule compiler, and forward chainer---is a faithful reproduction that handles all 30 IMO-AG-30 problems correctly. The 102-test suite provides confidence in the symbolic foundation.''
\end{sourcebox}
Handling the input representation and passing local tests are compatible with solving none of the benchmark problems, but ``handles ... correctly'' should not be read as end-to-end proof success. Likewise, the observed 0/30 is a result for this implementation and configuration; it does not establish a general impossibility result for forward chaining. The paper's appeal to the upstream system's performance does not validate the completeness of its own reproduction.

Most revealingly, the manuscript's ``Next Steps (Stage 1--5)'' section places the planned core components after the delivered study: ``Synthetic Data Generation'', ``GNN Architecture and Training'', ``MCTS Integration'', ``Self-Play Expert Iteration'', and ``Evaluation''. The proposed MCTS success criterion is to solve at least three of the 30 problems within 10,000 simulations. This is a future target, not an achieved result. The paper therefore corroborates the trace more strongly than a low score alone: the planned learning-and-search system was explicitly deferred while the foundation was packaged as the final contribution.

RAC's local contracts did capture real work---parser repairs, tests, and diagnostic artifacts---but their supported verdicts did not keep the missing stages active as prerequisites for final delivery. The negative case is consequently about the granularity and coverage of coordination obligations, not the uselessness of the intermediate work. A polished account of an incomplete system can satisfy local delivery checks while losing external credit for the actual research objective.

\paragraph{Cross-case synthesis.}
The two ARK trajectories illustrate altered corrective handoffs that produced observable repairs. EvoScientist on \mbox{Math-001} shows how delivery verification can distinguish a missing implementation from a nonempty response without changing the next agent. Agent Laboratory on \mbox{Math-003} shows greater task relevance under communication-enabled fixed-order execution, with weaker evidence for the communication mechanism itself. The final case bounds these benefits: artifact-level success must remain connected to the unfinished research objective. Together, the cases locate RAC's potential value at the interfaces between agents---what evidence is handed over, what obligation is assigned, and what work follows---rather than in the number of agents invoked or the presence of a positive local verdict.

% Compact alternative to appendix_all_hops.tex; do not replace the raw file.
% Include in the existing paper with \input{appendix_all_hops_compact}.
% Required preamble packages: longtable, booktabs, array, xcolor,
% and \usepackage[most]{tcolorbox}. These are already loaded in the main paper.
% Labels and environment names are distinct from those in the raw appendix.
\clearpage
\section{Condensed All-Hop Trajectories}
\label{app:all_hops_compact}
We summarize all 92 exported hops from \mbox{Physics-002} / ARK / R3 and \mbox{Life-001} / ARK / R2, retaining one row per hop and the chronological order. The tables condense reported work, corrective handoffs, and repeated checking; selected complete exported summaries follow. Phase headings are editorial groupings, not additional workflow stages.

\paragraph{Reading the tables.}
Work descriptions paraphrase agent-reported actions; they do not independently validate scientific results. In the final column, \emph{recorded next} is the successor specified by the host's native workflow, whereas \emph{observed next} is the agent selected by RAC and actually invoked next. \textcolor[HTML]{17365D}{\textbf{Blue entries}} highlight departures from the native successor; the record alone does not establish the rationale for each selection. Res., Exp., Wri., Rev., Plan., and Code denote researcher, experimenter, writer, reviewer, planner, and coder. A dash means no subsequent hop is exported. The host reviewer is distinct from the RAC verifier and the external benchmark judge.

These are qualitative records, not isolated estimates of causal benefit or matched-budget efficiency. The boxes following each table preserve selected complete exported summaries. Internal ratings are not external benchmark scores. Markdown syntax is retained as text, with typography and emoji markers adapted for print.

\begingroup
\definecolor{compactHopBlue}{HTML}{17365D}
\definecolor{compactHopGold}{HTML}{B99545}
\newtcolorbox{compacthopquote}[1]{enhanced,breakable,
  colback=compactHopBlue!3!white,colframe=compactHopGold,
  colbacktitle=compactHopBlue,coltitle=white,title={#1},
  fonttitle=\bfseries,fontupper=\small,boxrule=0.6pt,arc=2pt,
  left=7pt,right=7pt,top=5pt,bottom=5pt}
\setlength{\tabcolsep}{3pt}
\setlength{\LTleft}{0pt}
\setlength{\LTright}{0pt}
\renewcommand{\arraystretch}{1.16}

\subsection{\mbox{Physics-002} / ARK / R3}
\label{app:compact_physics}
The trajectory alternated textual revisions with coder and experimenter checks, including a numerical defect introduced in a later figure. The final benchmark score was 34.60; the archive stopped at the hard hop limit. Hops 3, 12, and 44 have no summary.

{\small
\begin{longtable}{@{}p{0.045\linewidth}>{\raggedright\arraybackslash}p{0.145\linewidth}>{\raggedright\arraybackslash}p{\dimexpr0.64\linewidth-6\tabcolsep\relax}>{\raggedright\arraybackslash}p{0.17\linewidth}@{}}
\caption{\mbox{Physics-002}: all 46 exported hops, summarized chronologically. Work and checks are agent-reported.}\label{tab:compact_physics}\\
\toprule
Hop & Agent & Reported work and outcome & Recorded $\to$ observed next \\
\midrule
\endfirsthead
\multicolumn{4}{@{}l}{\tablename~\thetable\ (continued)}\\
\toprule
Hop & Agent & Reported work and outcome & Recorded $\to$ observed next \\
\midrule
\endhead
\midrule
\multicolumn{4}{r@{}}{\footnotesize Continued on the next page}\\
\endfoot
\bottomrule
\endlastfoot
\addlinespace[5pt]
\multicolumn{4}{@{}l}{\textcolor{compactHopBlue}{\textbf{Initial research and first refinement (hops 0--6)}}}\\*
\addlinespace[2pt]
0 & Researcher & Established the RCS fidelity-estimation project context. & Exp. $\to$ Exp. \\
\addlinespace[2pt]
1 & Experimenter & Executed the initial five experiments and summarized results. & Wri. $\to$ Wri. \\
\addlinespace[2pt]
2 & Writer & Wrote and compiled the initial manuscript. & Rev. $\to$ Rev. \\
\addlinespace[2pt]
3 & Reviewer & No summary available; reviewer invocation retained. & \textcolor{compactHopBlue}{\textbf{Rev.} $\to$ \textbf{Plan.}} \\
\addlinespace[2pt]
4 & Planner & Prepared an action plan from the available review state. & Exp. $\to$ Exp. \\
\addlinespace[2pt]
5 & Experimenter & Checked the pipeline and extended experimental evidence. & Wri. $\to$ Wri. \\
\addlinespace[2pt]
6 & Writer & Revised the manuscript against the action plan. & Rev. $\to$ Rev. \\
\addlinespace[2pt]
\addlinespace[5pt]
\multicolumn{4}{@{}l}{\textcolor{compactHopBlue}{\textbf{Cross-agent numerical repairs (hops 7--15)}}}\\*
\addlinespace[2pt]
7 & Reviewer & Identified inconsistent XEB values and missing numerical evidence. & \textcolor{compactHopBlue}{\textbf{Plan.} $\to$ \textbf{Wri.}} \\
\addlinespace[2pt]
8 & Writer & Repaired tables, shot-count disclosure and figure caption. & \textcolor{compactHopBlue}{\textbf{Rev.} $\to$ \textbf{Code}} \\
\addlinespace[2pt]
9 & Coder & Cross-checked XEB values and corrected stale significance wording from 26 to 23 standard errors. & \textcolor{compactHopBlue}{\textbf{Wri.} $\to$ \textbf{Exp.}} \\
\addlinespace[2pt]
10 & Experimenter & Checked numerical artifacts, corrected Transport confidence intervals, and added the N56 Transport table. & Wri. $\to$ Wri. \\
\addlinespace[2pt]
11 & Writer & Added discussion of XEB/MB divergence. & Rev. $\to$ Rev. \\
\addlinespace[2pt]
12 & Reviewer & No summary available; reviewer invocation retained. & \textcolor{compactHopBlue}{\textbf{Rev.} $\to$ \textbf{Plan.}} \\
\addlinespace[2pt]
13 & Planner & Prepared the next correction and verification plan. & Exp. $\to$ Exp. \\
\addlinespace[2pt]
14 & Experimenter & Re-ran the pipeline and cross-checked six tables. & Wri. $\to$ Wri. \\
\addlinespace[2pt]
15 & Writer & Hedged residual confirmation claims and checked presentation. & Rev. $\to$ Rev. \\
\addlinespace[2pt]
\addlinespace[5pt]
\multicolumn{4}{@{}l}{\textcolor{compactHopBlue}{\textbf{Disclosure and interpretation revisions (hops 16--28)}}}\\*
\addlinespace[2pt]
16 & Reviewer & Reviewed repaired evidence and remaining disclosure issues. & \textcolor{compactHopBlue}{\textbf{Plan.} $\to$ \textbf{Wri.}} \\
\addlinespace[2pt]
17 & Writer & Added disclosure, significance method and result-file paths. & \textcolor{compactHopBlue}{\textbf{Rev.} $\to$ \textbf{Plan.}} \\
\addlinespace[2pt]
18 & Planner & Planned follow-up work after writer corrections. & \textcolor{compactHopBlue}{\textbf{Exp.} $\to$ \textbf{Code}} \\
\addlinespace[2pt]
19 & Coder & Checked prior presentation fixes and clarified the interpretation of the significance range. & \textcolor{compactHopBlue}{\textbf{Wri.} $\to$ \textbf{Exp.}} \\
\addlinespace[2pt]
20 & Experimenter & Checked the N56 MB table against experiment outputs. & Wri. $\to$ Wri. \\
\addlinespace[2pt]
21 & Writer & Checked prior fixes and updated the manuscript. & Rev. $\to$ Rev. \\
\addlinespace[2pt]
22 & Reviewer & Reviewed the paper and remaining minor issues. & \textcolor{compactHopBlue}{\textbf{Plan.} $\to$ \textbf{Wri.}} \\
\addlinespace[2pt]
23 & Writer & Clarified Transport wording, novelty and significance values. & \textcolor{compactHopBlue}{\textbf{Rev.} $\to$ \textbf{Plan.}} \\
\addlinespace[2pt]
24 & Planner & Prepared the next plan despite only line-level issues. & Exp. $\to$ Exp. \\
\addlinespace[2pt]
25 & Experimenter & Quantified the Transport hit-rate distribution and applied fixes. & Wri. $\to$ Wri. \\
\addlinespace[2pt]
26 & Writer & Checked experimenter edits and resolved remaining plan items. & Rev. $\to$ Rev. \\
\addlinespace[2pt]
27 & Reviewer & Reviewed resolved issues and remaining interpretation claims. & \textcolor{compactHopBlue}{\textbf{Plan.} $\to$ \textbf{Wri.}} \\
\addlinespace[2pt]
28 & Writer & Separated monotonic from exponential decay and qualified language. & \textcolor{compactHopBlue}{\textbf{Rev.} $\to$ \textbf{Plan.}} \\
\addlinespace[2pt]
\addlinespace[5pt]
\multicolumn{4}{@{}l}{\textcolor{compactHopBlue}{\textbf{Interval methods and statistical claims (hops 29--38)}}}\\*
\addlinespace[2pt]
29 & Planner & Recorded an action plan and protocol-coverage audit. & \textcolor{compactHopBlue}{\textbf{Exp.} $\to$ \textbf{Code}} \\
\addlinespace[2pt]
30 & Coder & Changed plotted intervals to pooled Clopper--Pearson intervals and updated captions and evidence records. & \textcolor{compactHopBlue}{\textbf{Wri.} $\to$ \textbf{Exp.}} \\
\addlinespace[2pt]
31 & Experimenter & Re-ran experiments and verified the implemented corrections. & Wri. $\to$ Wri. \\
\addlinespace[2pt]
32 & Writer & Checked all five issues as resolved on disk. & Rev. $\to$ Rev. \\
\addlinespace[2pt]
33 & Reviewer & Found a false within-one-standard-error agreement claim: the reported $N=40$ discrepancy was $1.83\sigma$. & \textcolor{compactHopBlue}{\textbf{Plan.} $\to$ \textbf{Wri.}} \\
\addlinespace[2pt]
34 & Writer & Replaced that claim with overlapping 95\% confidence intervals and explicitly reported the $1.83\sigma$ discrepancy. & \textcolor{compactHopBlue}{\textbf{Rev.} $\to$ \textbf{Plan.}} \\
\addlinespace[2pt]
35 & Planner & Planned follow-up work with the main issue already corrected. & \textcolor{compactHopBlue}{\textbf{Exp.} $\to$ \textbf{Code}} \\
\addlinespace[2pt]
36 & Coder & Checked earlier figure, execution and framing repairs. & \textcolor{compactHopBlue}{\textbf{Wri.} $\to$ \textbf{Exp.}} \\
\addlinespace[2pt]
37 & Experimenter & Regenerated page images and added a significance-metric caution. & Wri. $\to$ Wri. \\
\addlinespace[2pt]
38 & Writer & Checked the iteration-9 corrections in the manuscript. & Rev. $\to$ Rev. \\
\addlinespace[2pt]
\addlinespace[5pt]
\multicolumn{4}{@{}l}{\textcolor{compactHopBlue}{\textbf{Final checks, new figure, and residual repair (hops 39--45)}}}\\*
\addlinespace[2pt]
39 & Reviewer & Reviewed the manuscript and identified remaining consistency issues. & \textcolor{compactHopBlue}{\textbf{Plan.} $\to$ \textbf{Wri.}} \\
\addlinespace[2pt]
40 & Writer & Hedged the abstract Transport claim and revised presentation. & \textcolor{compactHopBlue}{\textbf{Rev.} $\to$ \textbf{Plan.}} \\
\addlinespace[2pt]
41 & Planner & Reported zero outstanding issues; the next observed agent was still a coder. & \textcolor{compactHopBlue}{\textbf{Exp.} $\to$ \textbf{Code}} \\
\addlinespace[2pt]
42 & Coder & Added an N56 depth-scan panel and qualified the Transport claim. & \textcolor{compactHopBlue}{\textbf{Wri.} $\to$ \textbf{Exp.}} \\
\addlinespace[2pt]
43 & Experimenter & Found a hardcoded shot-count denominator in the new panel; reported replacing it with actual totals. & Wri. $\to$ Wri. \\
\addlinespace[2pt]
44 & Writer & No summary available; the recorded error was \texttt{qualification\_\allowbreak request\_\allowbreak limit}. & Wri. $\to$ Wri. \\
\addlinespace[2pt]
45 & Writer & Reported prior fixes verified and corrected full-width figure layout; no later hop is exported. & Rev. $\to$ -- \\
\addlinespace[2pt]
\end{longtable}
}

\medskip
Hops 9--10 document consecutive coder and experimenter checks after the writer's revision. Hop 43 reports a defect in a newly added figure. The exported summaries for hops 10 and 43 may be truncated upstream at 2,000 characters; their available text is reproduced without reconstructing the missing tail.

\begin{compacthopquote}{Hop 9 \textperiodcentered{} coder}
\small
\textbf{Task /\allowbreak{} reported work}: Check the writer repairs and correct stale significance wording. Editorial heading inferred from the exported summary, not an assignment quote.\par\smallskip
\textbf{Handoff}: Recorded next: writer; observed next at hop 10: experimenter. The two fields differ.\par\smallskip
\textbf{Artifact changes}: Exported scientific\_\allowbreak{}changes = 3. No file-level diff is provided for this hop.\par\smallskip
\textbf{Recorded status}: done = false; error = null.\par\smallskip
\textbf{Complete available summary}\par\smallskip
{\footnotesize\ttfamily\raggedright\sloppy
\#\# Summary\par
\smallskip
I reviewed the state of `report/\allowbreak{}main.tex` against the reviewer's four acceptance conditions. The previous writer agent had already addressed all major issues (M1\textendash{}M5), but I discovered and fixed one residual error.\par
\smallskip
\#\#\# Change Made\par
\smallskip
**`report/\allowbreak{}main.tex` \textemdash{} Conclusion (\S{}6), line 234**: Corrected the d=24 N\ensuremath{\sigma} claim from **26\ensuremath{\sigma} \ensuremath{\rightarrow} 23\ensuremath{\sigma}**. The actual computed value from `outputs/\allowbreak{}n56\_\allowbreak{}mb\_\allowbreak{}transport\_\allowbreak{}depth\_\allowbreak{}scan\_\allowbreak{}instances.jsonl` is 23.0\ensuremath{\sigma} (mean=0.1208, SE=0.005244). The Discussion section already correctly stated "23\ensuremath{\sigma}" \textemdash{} the Conclusion was the only place with the stale "26\ensuremath{\sigma}".\par
\smallskip
\#\#\# Verification Results\par
\smallskip
| Acceptance Condition | Status | Evidence |\par
|---|---|---|\par
| (1) XEB table values reconciled | [check] | All 7 depths match JSONL data (d=8:0.632, d=10:0.450, d=12:0.457, d=14:0.397, d=16:0.368, d=18:0.266, d=20:0.302) |\par
| (2) N\ensuremath{\sigma} claims corrected | [check] | d=12: 26\ensuremath{\sigma}, d=24: 23\ensuremath{\sigma} (Discussion + Conclusion consistent) |\par
| (3) Figure 1 caption fixed | [check] | "plotted on the same [0,1] axis for compactness" |\par
| (4) N56 MB table added | [check] | Appendix table with 8 depths, CIs, shot counts |\par
\smallskip
LaTeX compiles clean (10 pages, 0 errors).\par
}
\end{compacthopquote}

\begin{compacthopquote}{Hop 10 \textperiodcentered{} experimenter}
\small
\textbf{Task /\allowbreak{} reported work}: Cross-check numerical artifacts and correct confidence intervals. Editorial heading inferred from the exported summary, not an assignment quote.\par\smallskip
\textbf{Handoff}: Recorded next: writer; observed next at hop 11: writer. The two fields agree.\par\smallskip
\textbf{Artifact changes}: Exported scientific\_\allowbreak{}changes = 6. No file-level diff is provided for this hop.\par\smallskip
\textbf{Recorded status}: done = false; error = null.\par\smallskip
\textbf{Complete available summary}\par\smallskip
{\footnotesize\ttfamily\raggedright\sloppy
\#\# Completed: Data Integrity Corrections for RCS Paper (Iteration 3 Review)\par
\smallskip
\#\#\# Summary of Changes\par
\smallskip
All reviewer-identified issues resolved. Paper now reconciles with experiment data at every claim.\par
\smallskip
\#\#\# Changes Made to `report/\allowbreak{}main.tex`\par
\smallskip
| Issue | Fix |\par
|---|---|\par
| **M1**: XEB table values | All 7 depths verified within \ensuremath{\pm}0.0005 of JSONL data \textemdash{} already correct |\par
| **M2**: N\ensuremath{\sigma} claims | d=12: 26\ensuremath{\sigma}, d=24: 23\ensuremath{\sigma} (both Discussion and Conclusion \textemdash{} already correct from previous iteration) |\par
| **M3**: Figure caption | "same \$[0,1]\$ axis for compactness" \textemdash{} already correct |\par
| **M4**: N56 MB table | Already present in Appendix (Table B) |\par
| **M5**: Shot-count inconsistency | Disclosed in \S{}3.1 and appendix table caption (d=24 uses 100 shots vs 20 elsewhere) |\par
| **m1**: Monotonicity claim | Text now acknowledges d=18\ensuremath{\rightarrow}20 uptick (0.266\ensuremath{\rightarrow}0.302), notes overlapping CIs indicate statistical fluctuation |\par
| **m2**: Transport CI values | Table 2 corrected: N=16 CI [0.946,1.000]\ensuremath{\rightarrow}[0.967,1.000]; N=24 [0.812,0.944]\ensuremath{\rightarrow}[0.867,0.913]; N=32 [0.800,0.936]\ensuremath{\rightarrow}[0.814,0.946]; N=40 [0.664,0.840]\ensuremath{\rightarrow}[0.663,0.857]; N=48 [0.664,0.840]\ensuremath{\rightarrow}[0.658,0.862]; N=56 [0.559,0.752]\ensuremath{\rightarrow}[0.520,0.800] |\par
| **m4**: N56 Transport table | Added Appendix Table D with 6 depths, Clopper-Pearson CIs |\par
| **m6**: N40\ensuremath{\rightarrow}N56 gap | Exact per-depth gaps now listed, d=16 anomaly (gap=0.057) noted, SPAM vs gate-error interpretation added |\par
\smallskip
\#\#\# Verifiable Evidence\par
\smallskip
- `report/\allowbreak{}main.tex` \textemdash{} corrected LaTeX source (line 129: monotonicity fix; lines 185-190: Transport CI fixes; lines 296-317: N56 Transport table; line 219: gap discussion)\par
- `report/\allowbreak{}main.pdf` \textemdash{} compiles clean (11 pages, 0 errors)\par
- `auto\_\allowbreak{}research/\allowbreak{}state/\allowbreak{}findings.json` \textemdash{} updated with review verification finding (f0\_\allowbreak{}review\_\allowbreak{}verification) and coverage entry\par
\smallskip
\#\#\# Correctness Audit Results\par
\smallskip
- XEB depth scan: max delta 0.00028 (within rounding)\par
- N\ensuremath{\sigma}: d=12 = 25.9\ensuremath{\sigma} \ensuremath{\approx} 26\ensuremath{\sigma} [check], d=24 = 23.0\ensuremath{\sigma} [check]\par
- Transport CIs: recomputed with t(df=9), truncated to [0,1]\par
- N40-N56 gaps: 0.111, 0.109, 0.089, 0.087, 0.057, 0\par
}
\end{compacthopquote}

\begin{compacthopquote}{Hop 43 \textperiodcentered{} experimenter}
\small
\textbf{Task /\allowbreak{} reported work}: Check the new panel and repair its shot-count denominator. Editorial heading inferred from the exported summary, not an assignment quote.\par\smallskip
\textbf{Handoff}: Recorded next: writer; observed next at hop 44: writer. The two fields agree.\par\smallskip
\textbf{Artifact changes}: Exported scientific\_\allowbreak{}changes = 6. No file-level diff is provided for this hop.\par\smallskip
\textbf{Recorded status}: done = false; error = null.\par\smallskip
\textbf{Complete available summary}\par\smallskip
{\footnotesize\ttfamily\raggedright\sloppy
\#\# Experimenter Verification \textemdash{} Iteration 10 Review Fixes\par
\smallskip
\#\#\# Status: Both reviewer issues verified [check], plus one bug found and fixed [bug]\par
\smallskip
\#\#\# 1. `review:execution:982d33ab5e` (Hedge transport claim in Introduction) \textemdash{} **VERIFIED ON DISK**\par
\smallskip
The abstract (line 33) and Introduction (line 51) both use hedged language. The abstract correctly frames the DeCross analysis as "*relied primarily on XEB*" and describes this paper's contribution as extending with three complementary methods \textemdash{} no claim that alternatives are "unavailable." The Introduction (line 51) reads: "*evidence that ion transport does not introduce a dominant transport-specific error source beyond standard decoherence scaling, though a small (5--10\%) contribution cannot be excluded at the current statistical resolution.*"\par
\smallskip
\#\#\# 2. `review:figure:b72031b754` (Add N56 MB panel to summary figure) \textemdash{} **VERIFIED ON DISK**\par
\smallskip
Panel (c) exists in `outputs/\allowbreak{}summary\_\allowbreak{}figure.pdf` showing N=56 MB depth scan (d=8\textendash{}24) with Transport 1QRB overlay (d=4\textendash{}96). Figure caption describes panel (c) with correct CI methodology. Body text (\S{}4.2) references panel (c).\par
\smallskip
\#\#\# 3. Data-Integrity Bug Found and Fixed [red marker]\par
\smallskip
**Root cause**: The panel (c) code hardcoded `n\_\allowbreak{}total = n * 20` for computing pooled Clopper-Pearson CIs, assuming all MB instances have 20 shots. However, **d=24 has 100 shots per instance** (50 instances \ensuremath{\times} 100 = 5000 trials). This inflated the d=24 CI by \textasciitilde{}30\%.\par
\smallskip
| | Buggy | Correct (fixed) | Appendix B Table |\par
|---|---|---|---|\par
| d=24 MB CI low | 0.1014 | 0.1119 | 0.112 |\par
| d=24 MB CI high | 0.1428 | 0.1302 | 0.130 |\par
\smallskip
**Fix**: Rewrote the panel (c) data loader to accumulate actual `n\_\allowbreak{}ideal` (successes) and `n\_\allowbreak{}total` (trials) from each JSONL row, then compute CP CIs from the actual totals. Same fix applied to Transport for robustness. Code changes in `code/\allowbreak{}run\_\allowbreak{}experiments.py` lines 801\textendash{}846.\par
\smallskip
\#\#\# Score adjustment: 8.0 \ensuremath{\rightarrow} 6.9/\allowbreak{}10\par
\smallskip
The hardcoded-shot-count bug is a data-integrity issue \textemdash{} d=24 in panel (c) is the single most important poi\par
}
\end{compacthopquote}

\subsection{\mbox{Life-001} / ARK / R2}
\label{app:compact_life}
The early trajectory repeatedly passed corrective work from the reviewer directly to the writer. Later hops returned to planning and repeated checks, including checks after a positive internal review. The final benchmark score was 14.10; the archive stopped at the hard hop limit. 
{\small
\begin{longtable}{@{}p{0.045\linewidth}>{\raggedright\arraybackslash}p{0.145\linewidth}>{\raggedright\arraybackslash}p{\dimexpr0.64\linewidth-6\tabcolsep\relax}>{\raggedright\arraybackslash}p{0.17\linewidth}@{}}
\caption{\mbox{Life-001}: all 46 exported hops, summarized chronologically. Work and checks are agent-reported.}\label{tab:compact_life}\\
\toprule
Hop & Agent & Reported work and outcome & Recorded $\to$ observed next \\
\midrule
\endfirsthead
\multicolumn{4}{@{}l}{\tablename~\thetable\ (continued)}\\
\toprule
Hop & Agent & Reported work and outcome & Recorded $\to$ observed next \\
\midrule
\endhead
\midrule
\multicolumn{4}{r@{}}{\footnotesize Continued on the next page}\\
\endfoot
\bottomrule
\endlastfoot
\addlinespace[5pt]
\multicolumn{4}{@{}l}{\textcolor{compactHopBlue}{\textbf{Initial research and report (hops 0--2)}}}\\*
\addlinespace[2pt]
0 & Researcher & Established the vaccine-design project context. & Exp. $\to$ Exp. \\
\addlinespace[2pt]
1 & Experimenter & Ran five experiments; reported repairing pivot indexing and verifying 17 output files. & Wri. $\to$ Wri. \\
\addlinespace[2pt]
2 & Writer & Wrote and compiled the initial manuscript. & Rev. $\to$ Rev. \\
\addlinespace[2pt]
\addlinespace[5pt]
\multicolumn{4}{@{}l}{\textcolor{compactHopBlue}{\textbf{Direct reviewer--writer corrections (hops 3--20)}}}\\*
\addlinespace[2pt]
3 & Reviewer & Reviewed the method--result mismatch and missing evidence. & \textcolor{compactHopBlue}{\textbf{Plan.} $\to$ \textbf{Wri.}} \\
\addlinespace[2pt]
4 & Writer & Added the MinSum curve alongside the greedy baseline; revised efficacy scope and numerical claims. & Rev. $\to$ Rev. \\
\addlinespace[2pt]
5 & Reviewer & Reviewed the first corrections and remaining scope claims. & \textcolor{compactHopBlue}{\textbf{Plan.} $\to$ \textbf{Wri.}} \\
\addlinespace[2pt]
6 & Writer & Qualified feasibility and added replicate variability. & Rev. $\to$ Rev. \\
\addlinespace[2pt]
7 & Reviewer & Checked earlier corrections and identified further text issues. & \textcolor{compactHopBlue}{\textbf{Plan.} $\to$ \textbf{Wri.}} \\
\addlinespace[2pt]
8 & Writer & Softened runtime scaling claims and revised text. & Rev. $\to$ Rev. \\
\addlinespace[2pt]
9 & Reviewer & Reviewed quantitative claims against outputs. & \textcolor{compactHopBlue}{\textbf{Plan.} $\to$ \textbf{Wri.}} \\
\addlinespace[2pt]
10 & Writer & Corrected the count of instances and scoped the budget claim. & Rev. $\to$ Rev. \\
\addlinespace[2pt]
11 & Reviewer & Reviewed the revised manuscript and underlying cell data. & \textcolor{compactHopBlue}{\textbf{Plan.} $\to$ \textbf{Wri.}} \\
\addlinespace[2pt]
12 & Writer & Checked low-response cells against source data and replaced a contradicted explanation. & Rev. $\to$ Rev. \\
\addlinespace[2pt]
13 & Reviewer & Reviewed the remaining runtime claim. & \textcolor{compactHopBlue}{\textbf{Plan.} $\to$ \textbf{Wri.}} \\
\addlinespace[2pt]
14 & Writer & Corrected runtime from under one second to the measured range. & Rev. $\to$ Rev. \\
\addlinespace[2pt]
15 & Reviewer & Identified MinSum separability and other text issues. & \textcolor{compactHopBlue}{\textbf{Plan.} $\to$ \textbf{Wri.}} \\
\addlinespace[2pt]
16 & Writer & Acknowledged separable ranking and revised the method description. & Rev. $\to$ Rev. \\
\addlinespace[2pt]
17 & Reviewer & Reviewed separability disclosure and threshold interpretation. & \textcolor{compactHopBlue}{\textbf{Plan.} $\to$ \textbf{Wri.}} \\
\addlinespace[2pt]
18 & Writer & Corrected the threshold range from 0.5--0.9 to 0.5--0.7. & Rev. $\to$ Rev. \\
\addlinespace[2pt]
19 & Reviewer & Reviewed the threshold fix and remaining terminology. & \textcolor{compactHopBlue}{\textbf{Plan.} $\to$ \textbf{Wri.}} \\
\addlinespace[2pt]
20 & Writer & Resolved three text-level issues from review iteration 9. & Rev. $\to$ Rev. \\
\addlinespace[2pt]
\addlinespace[5pt]
\multicolumn{4}{@{}l}{\textcolor{compactHopBlue}{\textbf{Renewed planning and checks (hops 21--28)}}}\\*
\addlinespace[2pt]
21 & Reviewer & Reported no remaining issues and a borderline-accept host-review verdict; execution nevertheless continued. & Plan. $\to$ Plan. \\
\addlinespace[2pt]
22 & Planner & Prepared an action plan and audited protocol coverage. & Exp. $\to$ Exp. \\
\addlinespace[2pt]
23 & Experimenter & Audited findings against experiment outputs. & Wri. $\to$ Wri. \\
\addlinespace[2pt]
24 & Writer & Reported an empty remediation list and no file modifications after rechecking the paper. & Rev. $\to$ Rev. \\
\addlinespace[2pt]
25 & Reviewer & Reviewed the manuscript again and identified further qualifications. & Plan. $\to$ Plan. \\
\addlinespace[2pt]
26 & Planner & Prepared the next action plan. & Exp. $\to$ Exp. \\
\addlinespace[2pt]
27 & Experimenter & Applied writing-only edits as the experimenter; reported running no new experiments. & Wri. $\to$ Wri. \\
\addlinespace[2pt]
28 & Writer & Checked corrections already made by the experimenter; reported no further writer edits were needed. & Rev. $\to$ Rev. \\
\addlinespace[2pt]
\addlinespace[5pt]
\multicolumn{4}{@{}l}{\textcolor{compactHopBlue}{\textbf{Further textual revisions and final review (hops 29--45)}}}\\*
\addlinespace[2pt]
29 & Reviewer & Reviewed the updated claims and comparisons. & \textcolor{compactHopBlue}{\textbf{Plan.} $\to$ \textbf{Wri.}} \\
\addlinespace[2pt]
30 & Writer & Qualified dominance language and applied three fixes. & Rev. $\to$ Rev. \\
\addlinespace[2pt]
31 & Reviewer & Reviewed title, claims and remaining issues. & \textcolor{compactHopBlue}{\textbf{Plan.} $\to$ \textbf{Wri.}} \\
\addlinespace[2pt]
32 & Writer & Resolved six issues including the coverage-guarantee title. & Rev. $\to$ Rev. \\
\addlinespace[2pt]
33 & Reviewer & Reviewed the next manuscript revision. & \textcolor{compactHopBlue}{\textbf{Plan.} $\to$ \textbf{Wri.}} \\
\addlinespace[2pt]
34 & Writer & Applied six review-requested corrections. & Rev. $\to$ Rev. \\
\addlinespace[2pt]
35 & Reviewer & Reviewed consistency after the corrections. & \textcolor{compactHopBlue}{\textbf{Plan.} $\to$ \textbf{Wri.}} \\
\addlinespace[2pt]
36 & Writer & Resolved instance labels and related textual issues. & Rev. $\to$ Rev. \\
\addlinespace[2pt]
37 & Reviewer & Checked prior fixes and flagged unsupported parameterization. & \textcolor{compactHopBlue}{\textbf{Plan.} $\to$ \textbf{Wri.}} \\
\addlinespace[2pt]
38 & Writer & Revised the unsupported instance-parameterization claim. & Rev. $\to$ Rev. \\
\addlinespace[2pt]
39 & Reviewer & Reviewed the revised manuscript and presentation. & \textcolor{compactHopBlue}{\textbf{Plan.} $\to$ \textbf{Wri.}} \\
\addlinespace[2pt]
40 & Writer & Resolved page-length and four further issues. & Rev. $\to$ Rev. \\
\addlinespace[2pt]
41 & Reviewer & Reviewed the revised runtime framing. & \textcolor{compactHopBlue}{\textbf{Plan.} $\to$ \textbf{Wri.}} \\
\addlinespace[2pt]
42 & Writer & Replaced linear-runtime claims with sub-quadratic wording. & Rev. $\to$ Rev. \\
\addlinespace[2pt]
43 & Reviewer & Reviewed remaining baseline and statistical descriptions. & \textcolor{compactHopBlue}{\textbf{Plan.} $\to$ \textbf{Wri.}} \\
\addlinespace[2pt]
44 & Writer & Added a justification for omitting top-k ranking, clarified statistical tests, and reduced abstract density. & Rev. $\to$ Rev. \\
\addlinespace[2pt]
45 & Reviewer & Confirmed prior fixes but flagged an unresolved abstract confidence-interval claim. & Plan. $\to$ -- \\
\addlinespace[2pt]
\end{longtable}
}

\medskip
Hop 12 reports replacing a speculative explanation with a data check. Hop 24 reports no remaining remediation items and no modifications, but the all-hop table shows that execution continued. Hop 45 records a remaining confidence-interval issue at the end of the exported trajectory.

\begin{compacthopquote}{Hop 12 \textperiodcentered{} writer}
\small
\textbf{Task /\allowbreak{} reported work}: Replace the low-response-cell speculation with a data check. Editorial heading inferred from the exported summary, not an assignment quote.\par\smallskip
\textbf{Handoff}: Recorded next: reviewer; observed next at hop 13: reviewer. The two fields agree.\par\smallskip
\textbf{Artifact changes}: Exported scientific\_\allowbreak{}changes = 4. No file-level diff is provided for this hop.\par\smallskip
\textbf{Recorded status}: done = false; error = null.\par\smallskip
\textbf{Complete available summary}\par\smallskip
{\footnotesize\ttfamily\raggedright\sloppy
The compilation confirms:\par
- Body ends at `[9]` (page 9), then `.bbl` loads\par
- 12 pages total (9 body + 3 appendix/\allowbreak{}references), clean, no errors\par
\smallskip
The fix is complete. Here's a summary of what was done:\par
\smallskip
---\par
\smallskip
\#\# Summary\par
\smallskip
**Task m1 \textemdash{} Fix E2 speculation at L147\textendash{}148**: The original text claimed the 8 low-response cells "likely represent a subclone whose presented peptides derive exclusively from a mutation not included in the vaccine." I cross-tabulated every low-response cell against its per-peptide scores in the raw data (`vaccine-elements.scores.100-cells.10x.rep-*.csv`) and found the claim was not just unsupported \textemdash{} it was **contradicted by the data**:\par
\smallskip
- **5 of 8 cells** have uniformly near-zero scores across all 12 candidates (p \ensuremath{\approx} 2\ensuremath{\times}10\textsuperscript{-}\textsuperscript{6} for every peptide), meaning no vaccine composition could cover them.\par
- **The remaining 3 cells** have their highest individual scores from **included** elements (mut26, mut44) \textemdash{} not excluded ones \textemdash{} but aggregate response still falls below 0.5.\par
\smallskip
**Fix applied**: Replaced the speculative claim with a data-grounded factual description at L147:\par
\smallskip
> "Of these 8 cells, 5 exhibit uniformly near-zero per-peptide response probabilities across all 12 candidates and would remain uncovered under any vaccine composition; the remaining 3 have moderate individual scores from included elements but insufficient aggregate response to clear the threshold."\par
\smallskip
All values are verified against the source CSVs. PDF compiles successfully (12 pages, body ends on page 9).\par
}
\end{compacthopquote}

\begin{compacthopquote}{Hop 24 \textperiodcentered{} writer}
\small
\textbf{Task /\allowbreak{} reported work}: Check a plan with no listed remediation issues. Editorial heading inferred from the exported summary, not an assignment quote.\par\smallskip
\textbf{Handoff}: Recorded next: reviewer; observed next at hop 25: reviewer. The two fields agree.\par\smallskip
\textbf{Artifact changes}: Exported scientific\_\allowbreak{}changes = 0. No file-level diff is provided for this hop.\par\smallskip
\textbf{Recorded status}: done = false; error = null.\par\smallskip
\textbf{Complete available summary}\par\smallskip
{\footnotesize\ttfamily\raggedright\sloppy
\#\# Summary\par
\smallskip
The action plan (`auto\_\allowbreak{}research/\allowbreak{}state/\allowbreak{}action\_\allowbreak{}plan.yaml`) explicitly states **`issues: []`** \textemdash{} no issues to remediate. The Iteration 10 reviewer gave a **Borderline Accept (7.5/\allowbreak{}10)** verdict with: "No further revisions needed. The manuscript is submission-ready."\par
\smallskip
I verified the paper is complete and correct:\par
\smallskip
\#\#\# Compilation\par
- **12 pages**, body ends on **page 8** (well within 9-page limit)\par
- Compiles cleanly with `pdflatex` + `bibtex` \textemdash{} no errors\par
- Only cosmetic underfull box warnings (typical for LaTeX)\par
\smallskip
\#\#\# Quantitative Claim Verification\par
Every number in the paper cross-checked against result files:\par
\smallskip
| Claim | Source | Verified |\par
|---|---|---|\par
| Greedy coverage: 55.7\% [52.6\%, 58.8\%] at k=1 | `coverage\_\allowbreak{}by\_\allowbreak{}budget\_\allowbreak{}aggregated.csv` | [check] |\par
| Greedy plateau: 96.4\% [95.1\%, 97.5\%] at k=6+ | `coverage\_\allowbreak{}by\_\allowbreak{}budget\_\allowbreak{}aggregated.csv` | [check] |\par
| MinSum k=4: 97.9\%, k=10: 99.2\% [98.4\%, 99.7\%] | `coverage\_\allowbreak{}by\_\allowbreak{}budget\_\allowbreak{}minsum.csv` | [check] |\par
| Mean P\_\allowbreak{}i = 0.943, median = 0.963, IQR = 0.047 | `cell\_\allowbreak{}response\_\allowbreak{}summary.json` | [check] |\par
| 99.2\% > 0.5, 88.7\% > 0.9 | `cell\_\allowbreak{}response\_\allowbreak{}summary.json` | [check] |\par
| 5 near-zero cells across reps 0, 1, 5, 8 | `cell\_\allowbreak{}response\_\allowbreak{}distribution.csv` | [check] |\par
| All 45 IoU = 1.0 | `vaccine\_\allowbreak{}stability\_\allowbreak{}iou.csv` | [check] |\par
| \ensuremath{\beta} range 0.99\textendash{}1.53, mean 1.23, all R\textsuperscript{2} \ensuremath{\geq} 0.94 | `runtime\_\allowbreak{}fit\_\allowbreak{}results.csv` | [check] |\par
| MinSum 99.2\% vs Greedy 96.4\% (5 wins, 5 ties) | `objective\_\allowbreak{}comparison.csv` | [check] |\par
| Random 91.9\%, p=0.002 Wilcoxon | `objective\_\allowbreak{}comparison.csv` | [check] |\par
\smallskip
\#\#\# Figure Placement\par
All 8 figures are present in `report/\allowbreak{}images/\allowbreak{}` and correctly referenced:\par
- Body: `fig\_\allowbreak{}coverage\_\allowbreak{}curve.pdf`, `fig\_\allowbreak{}response\_\allowbreak{}distribution.pdf`, `fig\_\allowbreak{}iou\_\allowbreak{}heatmap.pdf`, `fig\_\allowbreak{}runtime\_\allowbreak{}scaling.pdf`, `fig\_\allowbreak{}objective\_\allowbreak{}comparison.pdf`\par
- Appendix: `fig\_\allowbreak{}coverage\_\allowbreak{}curve\_\allowbreak{}per\_\allowbreak{}rep.pdf`, `fig\_\allowbreak{}response\_\allowbreak{}distribution\_\allowbreak{}per\_\allowbreak{}rep.pdf`, `fig\_\allowbreak{}runtime\_\allowbreak{}exponents.pdf`\par
\smallskip
\#\#\# No Modifications Made\par
The action plan has zero issues. The reviewer and planner both say the paper is submission-ready. No files were modified.\par
}
\end{compacthopquote}

\begin{compacthopquote}{Hop 45 \textperiodcentered{} reviewer}
\small
\textbf{Task /\allowbreak{} reported work}: Final recorded review: identify a remaining abstract CI ambiguity. Editorial heading inferred from the exported summary, not an assignment quote.\par\smallskip
\textbf{Handoff}: Recorded next: planner; no subsequent hop in this selected export. Do not infer execution of that successor.\par\smallskip
\textbf{Artifact changes}: Exported scientific\_\allowbreak{}changes = 0. No file-level diff is provided for this hop.\par\smallskip
\textbf{Recorded status}: done = false; error = null.\par\smallskip
\textbf{Complete available summary}\par\smallskip
{\footnotesize\ttfamily\raggedright\sloppy
\#\# Review Complete \textemdash{} Iteration 20\par
\smallskip
**Rating: Borderline Accept (7.3/\allowbreak{}10)**\par
\smallskip
\#\#\# Summary\par
\smallskip
I thoroughly reviewed all 12 PDF pages and the full LaTeX source. The paper has reached a high level of polish. All four issues from Iteration 19 are **resolved**:\par
\smallskip
- **M1**: Top-k baseline justification now lives in the E5 body (line 210) with explicit appendix cross-reference [check]\par
- **m1**: "near-linear" survivor at line 183 \ensuremath{\rightarrow} now reads "sub-quadratic" [check]\par
- **m2**: Implementation/\allowbreak{}Checklist now correctly describes mixed Wilcoxon/\allowbreak{}sign-test methodology (lines 121, 263, 267) [check]\par
- **m3**: Abstract numerical density reduced from \textasciitilde{}18 to \textasciitilde{}10 claims [check]\par
\smallskip
**One new minor issue identified (m1)**: The abstract's claim "95\% CI excludes \ensuremath{\beta}=1.0" is ambiguous. Per-instance CIs for instances 1 and 2 include 1.0, and the paper never computes a CI for the mean \ensuremath{\beta}. The body text correctly reports "95\% CIs exclude \ensuremath{\beta}=2.0 for six of seven instances" \textemdash{} the abstract should match this claim or compute the mean CI.\par
\smallskip
**Anonymity**: Clean \textemdash{} Anonymous Author(s)/\allowbreak{}Anonymous Institution, required ARK disclosure present, no identity leaks.\par
\smallskip
**Figures**: All clean, readable, properly scaled. Coverage curve, response distribution, IoU heatmap, runtime scaling, and objective comparison all present and well-formatted.\par
\smallskip
**Protocol alignment**: All 5 experiments delivered. Single-model limitation disclosed honestly. Findings.json surprises (perfect IoU, sub-quadratic runtime, single-population data) properly surfaced in the paper.\par
\smallskip
**No placeholder citations found** \textemdash{} all references appear to be real publications.\par
\smallskip
The review is saved to `auto\_\allowbreak{}research/\allowbreak{}state/\allowbreak{}latest\_\allowbreak{}review.md`.\par
}
\end{compacthopquote}

\endgroup

\section{Generative AI Use Disclosure}
\label{app:genai}

Generative AI systems are intrinsic experimental components of the evaluated AI-scientist workflows. They generate plans, code, experimental analyses, reviews, and manuscript drafts inside each evaluated system. Human authors are responsible for the research question, system design, evaluation protocol, verification of reported values, literature checking, interpretation, and final manuscript. All AI-assisted content was checked against persisted artifacts and source records before submission.

\end{document}